\documentclass[aps,pre,preprint,onecolumn]{revtex4-2}

\usepackage[T1]{fontenc}
\usepackage[utf8]{inputenc}
\usepackage{amssymb,amsmath}
\usepackage{graphicx}
\usepackage{xcolor}
\usepackage{siunitx}
\usepackage{mathrsfs}
\graphicspath{{Figures/}}
\usepackage{bm}
\usepackage{url}

\newcommand{\norm}[1]{\left\lVert#1\right\rVert}
\newcommand{\abs}[1]{\left|#1\right|}

\newcommand{\R}{\mathbb{R}}
\newcommand{\cont}[1]{\mathcal{C}^0\left(\left[#1\right]\right)}
\newcommand{\fck}[1]{\mathcal{C}^{k}\left(\left[#1\right]\right)}
\newcommand{\nmin}{N_{\text{min}}}
\newcommand{\dt}{\Delta t}
\DeclareMathOperator{\sgn}{sgn}

\newtheorem{prop}{Proposition}
\newtheorem{lem}{Lemma}
\newtheorem{cor}{Corollary}
\newtheorem{rem}{Remark}
\newenvironment{proof}[1][Proof]{\par\noindent\emph{#1.}\ }{\hfill$\square$\par}

\begin{document}

\title{Using waveform types, discrete Sobolev distances, and trend compatibility to cluster short time series of clinical data}

\author{Robyn J. N. Kettlitz}
\email{robynkettlitz@gmail.com}
\affiliation{Hangzhou International Innovation Institute, Beihang University, Hangzhou, China}
\author{Daniel Cisternas}
\affiliation{Escuela de Ingeniería, Universidad Mayor, Santiago, Chile}
\author{Alvaro Olivera-Nappa}
\affiliation{Centre for Biotechnology and Bioengineering -- CeBiB, Department of Chemical Engineering, Biotechnology and Materials -- DIQBM, University of Chile, Santiago, Chile}
\author{David Medina-Ortiz}
\email{david.medina@umag.cl}
\affiliation{Departamento de Ingeniería en Computación, Universidad de Magallanes, Punta Arenas, Chile}
\author{Seba Contreras}
\email{contreras@buaa.edu.cn}
\affiliation{Hangzhou International Innovation Institute, Beihang University, Hangzhou, China}
\affiliation{Interdisciplinary Center for the Mathematical Modeling of Infectious Disease Dynamics (IMMIDD), University of Münster, Münster, Germany}

\date{September 7, 2026}

%TC:ignore
\begin{abstract}
% background
Clinical tests, such as the oral glucose tolerance test (OGTT), perturb a physiological system at rest and record its excursion over a fixed protocol window as a short, undersampled time series.
% the problem
Such curves rarely allow parameter identification at the individual level, yet their trends carry physiological information that emerges at the population level as recurrent response patterns. Standard clustering methods aggregate differences globally and average out those trends, losing precisely the information about how the system regains control after the perturbation; identifying response patterns requires a method that preserves them.
% this work
We present a nonparametric framework that factorizes each curve into unit waveforms, sorts them with a symbolic trend-compatibility filter, and compares compatible shapes using discrete Sobolev-type distances on finite differences up to order $k$. The number of clusters, i.e., distinct response types, equals the number of realizable trend patterns the cohort populates, validated by resampling. Applied to 1283 five-point OGTTs, the method identifies 7 glycemia and 8 insulinemia types. The count holds across a threefold range of resolution. In benchmarks against six classical clustering methods, ours alone never mixes opposite trends, with comparable reconstruction error and stability. Clinical normality, assessed after clustering, is graded, overlapping, and distributed unevenly across response types ($h=0.14$--$0.48$).
% outlook
Overall, our method distinguishes response types among curves too short for alignment-based methods yet too informative for reduction to summary values. It applies to any protocol recording a system's response to a standardized perturbation on a short grid.
\end{abstract}

\maketitle
%TC:endignore

\clearpage
\section{Introduction}\label{sec:intro}

%Topic: Short time series arise are undersampled observations of a perturbed system returning to equilibrium, and although aren't enough to fit parameters, encode valuable information in their shape
Living organisms are complex adaptive systems: Their observable behavior results from the coordinated action of many interacting components. These components often compensate for one another, so the system can appear stable despite shifting internal regulation. For the observer, this compensation means that different internal states may produce nearly identical output trajectories \cite{edelman2001degeneracy}, making it difficult to infer mechanistic parameters from data \cite{gutenkunst2007universally,raue2009structural}. Clinical tests already reflect this complexity: they treat the organism as a dynamical system, apply a standardized perturbation to its equilibrium, and record the return trajectory. However, ethical, economic, and technical constraints often limit the number of measurements, so most clinical tests have short time series. The oral glucose tolerance test (OGTT) is a prime example: a fasting patient receives a glucose load, and glycemia and insulinemia are measured at a handful of time points over two hours. Standard practice reduces this trajectory to summary values (usually fasting and two-hour) or surrogate indices \cite{matsuda1999insulin,dallaman2002oral}, discarding most of the curve's shape, even though risk varies continuously within and across these categories \cite{decode2001glucose,oliveranappa2022patient}. Yet the shape itself carries early signatures of dysfunction \cite{tschritter2003shape,hulman2018glucose,contreras2020novel} and is directly measured. 

%Topic: Global distances and the available clustering methods lose the local trends that carry the information in short time series.
At the population level, identifying response types requires a distance metric that preserves meaningful features. Standard choices, such as the Euclidean, elastic, and kernel-based distances \cite{cuturi2017softdtw,marcoblanco2024kernels,chouakria2007cort}, aggregate differences globally, so that disagreement in a local trend can be hidden by agreement elsewhere. In short time series such as the five-point OGTT \cite{contreras2020novel,oliveranappa2022patient}, every local trend encodes important physiological information. These trends distinguish physiological regimes \cite{tschritter2003shape} that correspond to different levels of impairment in glycemic control. When local trends are averaged away, clusters lose interpretability and clinical value. Existing clustering methods are not designed for this situation: generic vector clustering treats the time points as unordered coordinates and represents no local transition \cite{saxena2017review}, and time-series methods typically require longer series for alignment or derivative-based comparisons \cite{keogh2001ddtw,gorecki2013derivatives,lafabregue2022deep,durso2000dissimilarity,durso2005fuzzy,gorecki2014second,villmann2007sobolev}. Major surveys omit short-curve settings \cite{paparrizos2024decade,holder2024elastic,aghabozorgi2015time}, and functional data analysis assumes a smooth latent curve \cite{wang2016functional,jacques2013funclust,yao2005sparse,yu2025derivative}. Latent-class models predefine admissible shapes \cite{nagin1999trajectories,twisk2012classifying,denteuling2021comparison,morris2013identification}, while other strategies focus on indices or continuous monitoring \cite{wagner2021pathophysiology,metwally2024cgm}. For $n=5$, latent-class model selection itself becomes unstable. Thus, key local trends are often lost through compression or omission.

%Topic: Symbolic encoding of the signs keeps the local trends.
If all distance metrics compress information, how can we retain the essential local trends? Symbolic dynamics offers a solution by encoding a signal by the order relations or discrete classes of its values \cite{kurths1995quantitative,bandt2002permutation}, which is particularly robust to noise \cite{zanin2012permutation,daw2003symbolic}. In certain systems, this order-based encoding does not lose information: for piecewise-monotone interval maps, permutation entropy equals Kolmogorov-Sinai entropy \cite{bandt2002interval}. By filtering on the signs of finite differences, we can group curves by compatible response shapes and compare only those with similar behavior. Clinical criteria are then read within each response shape; they play no role in defining it.

%Topic: STSC types short curves by waveform, and its new ingredient is the compatibility reading of the neutral symbol.
Here, we introduce a nonparametric framework for typing short clinical time series, i.e., undersampled excursions of a dynamical system after a controlled perturbation, observed over a standardized protocol window. We apply this method to a cohort of $1283$ patients who underwent an OGTT: a standardized $75$~g glucose load, with glycemia and insulinemia measured at $0$, $30$, $60$, $90$, and $120$~min \cite{jagannathan2020oral,contreras2020novel,oliveranappa2022patient}. Each curve is factorized into unit waveforms (Sec.~\ref{sec:distance}), which are compared using a family of discrete Sobolev-type distances $D_s^k$ based on finite differences up to order $k$, i.e., an extension of value-and-derivative methods \cite{durso2000dissimilarity,villmann2007sobolev}. Before comparison, curves are filtered based on their shape profiles and sorted into trend-compatibility classes (Sec.~\ref{sec:filter}), thereby enabling extension to arbitrary finite-difference and noise-tolerance orders. A four-stage algorithm (Sec.~\ref{sec:algorithm}) converts both representations into waveform types, with the number of types read from the occupancy of the realizable patterns. In sum, our approach translates five measurements into a classification of the glycemic response to the perturbation, thereby generalizing to any standardized excursion recorded on a short grid.

\section{Sobolev-type distances for short time series}\label{sec:distance}

%Topic: This section gives the mathematical backbone, and it applies only to series that are excursions from and back to an equilibrium.
This section provides the mathematical backbone of the proposed method, including the factorization and the definitions of the distances and filters that enable physiologically consistent identification of response types in short time series. These series must record the excursion of a dynamical system, in our case, the human physiological response, after a controlled perturbation from a resting state; the window need not contain the return to it.

%Topic: Every curve is factorized into level, amplitude, and a unit waveform.
The first step is factorizing the curves into waveforms, representing their dimensionless response as
\begin{equation}\label{eq:factorization}
x(t) = x_0 + A\,w(t),\qquad x_0 = x(0),\qquad A = \max_t\,\abs{x(t)-x_0},
\end{equation}
so that the waveform $w$ satisfies $w(0)=0$ and $\max_t\abs{w(t)}=1$. We naturally assume that none of the input curves is constant. The distance $D_s^k$ built below is evaluated on $w$, so curves are compared by shape alone.

\subsection{Continuous formulation of the distance}\label{sec:continuous}

%Topic: A weighted Sobolev-type norm induces a distance on smooth functions, with every order dimensionally consistent.
For $f\in\cont{t_0,t_{n-1}}$ and $1\leq p<\infty$, let
\begin{equation}\label{normap}
\norm{f}_p = \left(\int_{t_0}^{t_{n-1}}\abs{f\left(x\right)}^p\,dx\right)^{\frac{1}{p}}
\end{equation}
denote the $L^p$ norm. On the space $\fck{t_0,t_{n-1}}$ of $k$-times continuously differentiable functions we define, for fixed weights $\bm{\alpha}=(\alpha_0,\dots,\alpha_k)$ with $\alpha_i\geq 0$,
\begin{align}
    d_s^k &:\fck{t_0,t_{n-1}}\times\fck{t_0,t_{n-1}}\longrightarrow\R_{\geq 0},\label{def:dsk}\\
    d_s^k(f,g) & = \sum_{i=0}^k\alpha_i\norm{f^{(i)}-g^{(i)}}_p,\nonumber
\end{align}
where $f^{(i)}$ denotes the $i$-th derivative of $f$ ($f^{(0)} = f$). This is the distance induced by a weighted Sobolev-type norm, in the spirit of elastic shape analysis \cite{srivastava2011elastic}. Note that $\alpha_i$ has units of $x^{-1}\,\text{time}^{\,i-1/p}$.

\begin{prop}[Metric properties of $d_s^k$]\label{prop:metric}
If $\alpha_0>0$, then $d_s^k$ defines a metric on the space $\fck{t_0,t_{n-1}}$.
\end{prop}

\begin{proof}
Each summand in $d_s^k$ is nonnegative and symmetric in $(f,g)$. If $d_s^k(f,g) = 0$, then $\alpha_0 \norm{f-g}_p = 0$ with $\alpha_0 > 0$, so $f \equiv g$. The converse follows directly from the definition. Finally, writing $f^{(i)} - g^{(i)} = (f^{(i)} - h^{(i)}) + (h^{(i)} - g^{(i)})$, where $h\in \fck{t_0,t_{n-1}}$, and applying Minkowski's inequality to each term,
\begin{equation*}
    d_s^k(f,g) = \sum_{i=0}^k\alpha_i\norm{\big(f^{(i)}-h^{(i)}\big)+\big(h^{(i)}-g^{(i)}\big)}_p
    \leq d_s^k(f,h) + d_s^k(h,g).
\end{equation*}
\end{proof}

\begin{rem}[Pseudometric and shape-only comparison]\label{rem:pseudo}
Suppose $\bm\alpha \neq 0$ but $\alpha_0 = 0$, and let $j = \min\{i : \alpha_i > 0\}$. Then $d_s^k$ is a pseudometric: $d_s^k(f,g) = 0$ if and only if $f-g \in \mathcal{P}_{j-1}$, the space of polynomials of degree at most $j-1$ (the $i=j$ term forces $f^{(j)}\equiv g^{(j)}$, and integrating $j$ times gives the claim). Thus, $d_s^k$ becomes a metric on the quotient $\fck{t_0, t_{n-1}} / \mathcal{P}_{j-1}$. In particular, for $\alpha_0 = 0$, $\alpha_1 > 0$, functions differing by an additive constant are identified; this is the behavior desired when only the \emph{shape} of the response matters, playing the role of per-series standardization \cite{paparrizos2015kshape} without distorting higher-order structure.
\end{rem}

\subsection{Finite differences on coarse grids: exact mean-value structure}\label{sec:discrete}

%Topic: The data arrive on a grid of $n$ equally spaced timepoints.
Short time series such as the OGTT response curves sample $f$ at $n$ equally spaced timepoints $T_{\exp} = \{t_0, \dots, t_{n-1}\}$ with spacing $\Delta t$ (see Appendix~\ref{ap:uneven} for uneven grids). Given the discrete nature of the data, we use forward differences (the discrete analogs of derivatives), defined as
\begin{equation}\label{DD1}
\Delta^1f = f(t_{i+1})-f(t_{i}),\qquad
\Delta^mf = \sum_{j=0}^{m}\left(-1\right)^{j}\binom{m}{j}f\big(t_{i}+(m-j)\dt\big).
\end{equation}

%Topic: On a coarse grid the difference quotient never reaches the derivative, and the window must fit inside the curve.
Note that the limit $\Delta^m f/(\dt)^m \to f^{(m)}(t_i)$ is not reached on coarse grids, and $\Delta^m f$ is defined only for $i \leq n-1-m$, so that the window fits within the curve.

\begin{lem}[Mean-value representation of forward differences]\label{lem:mvt}
Let $m\geq 1$ and $f\in\mathcal{C}^{m}([t_i,t_{i+m}])$ on a uniform grid with step $\dt>0$. Then there exists $\xi\in(t_i,t_{i+m})$ such that
\begin{equation}\label{eq:mvt}
\Delta^mf = (\dt)^m\, f^{(m)}(\xi).
\end{equation}
\end{lem}

%Topic: The lemma is proved in the appendix, on standard results.
A proof is given in Appendix~\ref{ap:proofs}, following standard results \cite{deboor2005divided}.

\begin{cor}[Sign fidelity]\label{cor:sign}
Under the assumptions of Lemma~\ref{lem:mvt}:

\begin{enumerate}
    \item[(a)] There exists $\xi \in (t_i, t_{i+m})$ such that $\sgn(\Delta^m f) = \sgn(f^{(m)}(\xi))$. In other words, if $\Delta^m f > 0$ (or $< 0$, or $= 0$), then $f^{(m)}$ takes a positive (or negative, or zero) value somewhere in $ (t_i, t_{i+m}) $.
    \item[(b)] If $f^{(m)} > 0$ (or $< 0$) in $(t_i, t_{i+m})$, then $\Delta^m f > 0$ (or $< 0$).
\end{enumerate}

\end{cor}

%Topic: Part (b) licenses the sign filtering: a derivative of constant sign is reported faithfully at any grid coarseness.
Part (a) follows from Lemma~\ref{lem:mvt} divided by $(\dt)^m > 0$. Part (b) is what justifies the sign filtering of Sec.~\ref{sec:filter}: if the underlying derivative maintains a constant sign across the window, the finite difference reports it faithfully, regardless of grid coarseness.

\begin{rem}[Magnitudes are only first-order accurate]\label{rem:magnitude}
If moreover $f\in\mathcal{C}^{m+1}([t_i,t_{i+m}])$, then by Lemma~\ref{lem:mvt}, applying the mean value theorem to $f^{(m)}$ on $[t_i,\xi]$, and noting that $\sup_{[t_i,\xi]}\abs{f^{(m+1)}} \leq \sup_{[t_i,t_{i+m}]}\abs{f^{(m+1)}}$, we have
\begin{equation}\label{eq:magerror}
\abs{\frac{\Delta^mf}{(\dt)^m} - f^{(m)}(t_i)} \leq m\,\dt \sup_{[t_i,t_{i+m}]}\abs{f^{(m+1)}}.
\end{equation}
\end{rem}

%Topic: Magnitudes carry $O(\dt)$ error while signs stay robust, which is the asymmetry the filter exploits.
This yields an error of first order in $\dt$ for all $m$. Thus, derivative magnitudes estimated from coarse finite differences carry $O(\dt)$ uncertainty, yet their signs (under the window condition of Corollary~\ref{cor:sign}b and in the absence of noise) remain robust.

%Topic: The discrete distance $D_s^k$ is a metric whenever the value term is kept, and a pseudometric otherwise.
Scaling each difference by $(\dt)^{-i}$, the discrete version of the distance reads
\begin{align}\label{def:Dsk}
D_s^k &:\R^{n}\times\R^{n}\longrightarrow\R_{\geq 0},\qquad k\leq n-1,\\
D_s^k(\vec{x},\vec{y}) & = \sum_{i=0}^k\alpha_i\norm{\frac{\Delta^i[\vec x]}{(\dt)^i}-\frac{\Delta^i[\vec y]}{(\dt)^i}}_p,\nonumber
\end{align}
where $\Delta^i[\vec{x}] \in \R^{n-i}$ is the vector of $i$-th forward differences, and $\norm{\cdot}_p$ denotes the vector $p$-norm. Unless otherwise specified, we employ length-normalized norms $\norm{\vec{v}}_{p,i} = \left((n-i)^{-1} \sum_j |v_j|^p \right)^{1/p}$ in Eq.~\eqref{def:Dsk}, so that each order contributes by typical entry size rather than by vector length. Each map $\vec{x} \mapsto \norm{\Delta^i[\vec{x}] / (\dt)^i}_{p,i}$ is a seminorm on $\R^n$ and the $i=0$ one is positive definite, so Proposition~\ref{prop:metric} carries over: $D_s^k$ is a metric on $\R^n$ when $\alpha_0 > 0$. For $\alpha_0 = 0$, $D_s^k$ is a pseudometric with null space $\ker \Delta^j$, $j = \min\{i : \alpha_i > 0\}$, the grid-restricted analog of Remark~\ref{rem:pseudo} (Appendix~\ref{ap:proofs}).

\begin{rem}[Recovering the M\"oller-Levet distance]\label{rem:mollerlevet}
For $k=1$, $\alpha_0=0$, $\alpha_1=1$, and $p=2$ on a uniform grid, $D_s^1$ with the standard vector $2$-norm is the same as the short-time-series distance of Ref.~\cite{MOLLERLEVET200549}. With the length-normalized convention, it equals $(n-1)^{-1/2} d_{\mathrm{ML}}$. $d_{\mathrm{ML}}$ is a pseudometric, vanishing only when the two vectors differ by a constant (the $j=1$ case above). The family in Eq.~\eqref{def:Dsk} generalizes this to any order, includes the value term, and makes the weighting explicit.
\end{rem}

\subsection{Data-driven normalization and the choice of $k$}\label{sec:weights}

%Topic: Data-driven weights make the orders commensurate, at the cost of dataset dependence.
The orders in Eq.~\eqref{def:Dsk} span different scales: for raw glycemia curves, values are typically $O(100)$~\si{mg/dL}, while second differences are $O(10)$. If all $\alpha_i$ are one, the value term dominates and $D_s^k$ approaches the Euclidean distance; we therefore normalize each order by a robust scale from the empirical distribution of the dataset $Y=\{\vec x_1,\dots,\vec x_N\}$:
\begin{equation}\label{eq:weights}
\alpha_i = \frac{w_i}{s_i},\qquad
s_i = \operatorname{median}\Big\{\,\abs{\Delta^i[\vec x_r](t_j)/(\dt)^i}\;:\; r\leq N,\ j\leq n-1-i \,\Big\}.
\end{equation}
User weights $w_i \geq 0$ (default 1) allow emphasis on certain orders. If more than half of the pooled $i$-th differences vanish, $s_i = 0$ and the order is dropped ($w_i = 0$); dropping $i = 0$ returns the pseudometric case above. Equation~\eqref{eq:weights} is a per-feature standardization: each $s_i$ has units $x/(\text{time})^i$, so every term is dimensionless and commensurate. The median makes it robust to outliers and extreme values, and $D_s^k$ is invariant to a global rescaling $\vec x \mapsto c\,\vec x$, although not to translations ($s_0$ is a magnitude rather than a dispersion). The $s_i$ are computed once per dataset, so the metric properties are unchanged, at the cost of dataset dependence. In the case study, $D_s^k$ acts on the factorized waveform $w$ of Sec.~\ref{sec:distance}, and the scales $s_i$ are computed on $w$ itself.

%Topic: Three bounds on the difference order leave $k=1$ as the reference for the five-point OGTT.
Three considerations set an upper bound on $k$. First, a hard bound: $\Delta^k$ requires $k+1$ points, so $k \leq n-1$. Second, a two-window rule: order-$m$ differences only capture features slower than the window $m\,\Delta t$, and a feature of duration $T_{\min}$ should span at least two windows,
\begin{equation}\label{Shannon}
k\lesssim\frac{T_{\min}}{2\,\dt}.
\end{equation}
It limits $k$ by the shortest meaningful feature $T_{\min}$ and is a heuristic; the operational requirement is that $f^{(m)}$ keeps one sign across each window (Corollary~\ref{cor:sign}b). Third, noise sets another practical ceiling: for additive i.i.d. noise of variance $\sigma^2$, the variance of $\Delta^m/(\Delta t)^m$ scales as $\binom{2m}{m} \sigma^2 / (\Delta t)^{2m}$, which grows rapidly with $m$. For the five-point OGTT ($\Delta t=30$ min, relevant features $T \gtrsim 60$ min), the hard bound gives $k \leq 4$, the window condition $m \Delta t \leq T$ gives $k \leq 2$, and the two-window rule of Eq.~\eqref{Shannon} sets $k = 1$. The case study of Sec.~\ref{sec:casestudy} adopts $k = 1$ as its reference, for the rule of Eq.~\eqref{Shannon} and for the plateaus of the count reported in Sec.~\ref{sec:baselines}, and reports $k = 2$ as a refinement.

\section{Trend compatibility filter}\label{sec:filter}

%Topic: The resolution $\varepsilon$ decides which trends can be certified, and the neutral symbol stays compatible with either trend.
We call the sign of the $m$-th derivative over a window the order-$m$ \emph{trend} of a curve: rising or falling for $m=1$, accelerating or decelerating for $m=2$. Corollary~\ref{cor:sign} states that finite differences report these trends faithfully whenever the derivative keeps one sign across the window. Following this, we split the clustering problem into two stages: a \emph{trend compatibility filter}, which sorts curves by the trends we can certify they share, and a similarity quantification under $D_s^k$. The filter depends on a single parameter, the \emph{resolution} $\varepsilon>0$: the smallest change of the waveform per grid step that we accept as a real trend. Since $w$ is unit-normalized (Eq.~\eqref{eq:factorization}), $\varepsilon$ is a dimensionless fraction of the excursion. Differences smaller than the resolution may arise due to noise or to a sign change of $f^{(m)}$ within the window (Corollary~\ref{cor:sign}b); in that case, they receive the neutral symbol $0$. Explicitly, the trend of order $m$ at position $j$, i.e., the filtered sign of $\Delta^m_j[\vec w]$, the $j$-th entry of $\Delta^m[\vec w]$, reads
\begin{equation}\label{Filtro}
    \mathscr{F}\big(\Delta^m_j[\vec w]\big) = \left\{
	\begin{array}{ll}
		\sgn\big(\Delta^m_j[\vec w]\big) & \mbox{if } \abs{\Delta^m_j[\vec w]} \geq \varepsilon_m, \\[2pt]
		0 & \mbox{otherwise},
	\end{array}
\right.
\qquad
\varepsilon_m = \varepsilon\sqrt{\tbinom{2m}{m}\big/2},
\end{equation}
where the order-dependent scaling sets the tolerance to spurious trends. Additive noise of variance $\sigma^2$ on $\vec w$ induces variance $\binom{2m}{m}\sigma^2$ on $\Delta^m$ (Sec.~\ref{sec:weights}); to keep the probability of reading noise as a trend equal across orders, we rescale the threshold $\varepsilon_m$ at each order, with $\varepsilon_1=\varepsilon$. Concatenating the trends of orders $1,\dots,k$ defines the \emph{trend pattern} map
\begin{equation}\label{eq:pattern}
\pi:\R^n\longrightarrow\{-1,0,+1\}^{K},\qquad K=\sum_{m=1}^{k}(n-m)=kn-\frac{k(k+1)}{2},
\end{equation}
with $K=4$ local trends per curve at $k=1$ and $K=7$ at $k=2$ for $n=5$. The neutral symbol $0$ represents points where trends cannot be ascertained at resolution $\varepsilon$, and the curve can be assigned to either trend; we call this the \emph{compatibility reading} of the neutral symbol. Two curves are \emph{compatible} if and only if no position assigns them opposite trends. Compatibility is reflexive and symmetric. However, it is not transitive due to the flexible interpretation of the neutral symbols, and thus induces no partition. Section~\ref{sec:algorithm} builds the types by covering the cohort instead of partitioning it.

%Topic: The candidate types are the realizable fully specified trend patterns, $16$ of them at $k=1$ and $54$ at $k=2$.
The candidate types are the \emph{realizable fully specified trend patterns}, i.e., patterns without neutral symbols that at least one curve can produce. At $k=1$, every sequence of rising and falling steps is realizable, giving $2^{n-1}=16$ candidates for $n=5$. At $k=2$, slope and curvature trends are algebraically coupled, and only $2\cdot3^{n-2}=54$ assignments survive; this is the general-position row of the pattern count in Appendix~\ref{ap:count} (Proposition~\ref{prop:count}). In the unfiltered limit $\varepsilon\to0$ and at $m=1$, $\pi$ encodes the up/down sequence of the curve, which, on series without ties, coincides with the order-2 ordinal patterns of consecutive pairs \cite{bandt2002permutation}. Exact ties, which ordinal encodings break by convention, receive the third symbol $0$ here.

%Topic: Noise reaches the encoding only below the resolution; in measurement units the resolution is a per-curve quantity, and the reproducibility of the test decides which curves it resolves.
Measurement noise enters Eq.~\eqref{Filtro} only through sign flips of differences below the resolution, which the filter absorbs by construction. At the reference resolution $\varepsilon=0.1$, the filter neutralizes $12.9\%$ of the glucose and $18.1\%$ of the insulin first differences of the waveform; on glucose, the share increases along the curve, from $0.8\%$ at the first step to $12.3$, $17.6$ and $20.7\%$ at the following three. In measurement units, the resolution is $\varepsilon A$ and thus differs between curves: at $\varepsilon=0.1$, it is $5.7$~\si{mg/dL} for the median glucose curve, with $A$ ranging from $22$ to $119$~\si{mg/dL} between the 5th and 95th percentiles. The reproducibility of the test itself sets the noise the encoding must tolerate: repeated OGTTs in the same subjects show a within-subject standard deviation of the $2$-h glucose value near $0.9$~\si{mmol/L}, i.e., about $16$~\si{mg/dL} \cite{mooy1996intra}. That value exceeds $\varepsilon A$ for almost every curve, so the resolution cannot be set by it; we set it where the type count presents a plateau (Sec.~\ref{sec:baselines}), and Appendix~\ref{ap:sensitivity} reports the effect of noise of that size on the types and on the individual labels.

%Topic: STEM is the closest precursor, and our encoding differs from it in four aspects.
The closest precursor of our encoding is STEM \cite{ernst2005short}, which encodes short expression courses as up, down, and flat transitions, enumerates the admissible profiles, and assigns each curve to the nearest one. Ours differs in four aspects. First, our threshold is an absolute resolution on the unit waveform, whereas STEM fixes a transition level. Second, our code extends to arbitrary difference orders, thereby incorporating curvature trends in the description. Third, Lemma~\ref{lem:mvt} justifies why the trends, and not the magnitudes, are the observables to keep. Fourth, a trend pattern is a candidate type, not a final cluster; the metric of Sec.~\ref{sec:algorithm} afterward separates the curves compatible with it. Codes built on filtered trends are well established \cite{phang2003trajectory,kim2007dibc,peddada2003gene,brandmaier2015pdc}; to our knowledge, none of them combines higher difference orders with the compatibility reading on a protocol grid.

\section{The clustering algorithm}\label{sec:algorithm}

%Topic: The algorithm takes five parameters and returns the number of types as an output.
We designed our method to discover response types in a population; the number of types is an output of the algorithm. The clustering depends on three parameters, the difference order $k$, the resolution $\varepsilon$, and the minimal occupancy $\nmin$; the resampling certificate adds two more, $B$ and $\phi$ (Table~\ref{tab:params}). The algorithm then proceeds in four stages.

%Topic: Stage 1 factorizes every curve and hands the waveform to the later stages.
\emph{Stage 1, factorize.} Every curve is factorized into its waveform $w$ by Eq.~\eqref{eq:factorization}; every later stage acts on $w$ alone.

%Topic: Stage 2 encodes each waveform into its trend pattern and fixes the candidate types, without computing distances.
\emph{Stage 2, encode.} Each waveform is encoded by Eq.~\eqref{Filtro} at resolution $\varepsilon$ into its trend pattern $\pi(\vec w)$ of Eq.~\eqref{eq:pattern}, and the candidate types are the realizable fully specified patterns of Sec.~\ref{sec:filter}. Encoding is $O(Nnk)$ in complexity and computes no distances.

%Topic: Stage 3 chooses the types by a greedy cover and assigns the compatible curves by the metric.
\emph{Stage 3, cover and assign.} A greedy cover chooses the types: at each step, it adds the candidate compatible with the most not-yet-covered curves, each curve weighted by the reciprocal of the number of candidates it is compatible with. It stops when the candidate so selected would newly cover fewer than $\nmin$ curves, so the floor acts on how many realizable patterns the cohort populates. Curves compatible with several of the chosen types are then resolved by the metric: each joins the nearest compatible type, where nearest means the smallest $D_s^k$ on $w$ to the type medoid. Medoids are recomputed from the assignments until the labels stop changing; convergence is observed rather than guaranteed, and the implementation stops after $20$ iterations. Ties are broken deterministically: the cover takes the lowest pattern index, the medoid the lowest curve index, and the nearest type the lowest type index. A type left with fewer than $\nmin$ members is dropped, and its members are reassigned. Curves that are compatible with no chosen type are separated and reported rather than forced into a type.

%Topic: Stage 4 resamples to test whether the count and the labels survive; the Monte Carlo never selects the count.
\emph{Stage 4, resample.} Stages 1--3 are rerun on $B$ subsamples of fraction $\phi$, with the scales $s_i$ of Eq.~\eqref{eq:weights} computed once on the full cohort. The replicates return five quantities: the mode of the type count, pattern- and medoid-matched label alignments, plurality labels, per-curve residual frequencies, and pairwise ARI stability. We report the modal count and the plurality labels; the full-cohort run of Stage 3 is kept as a reference, and at the reference setting $98$--$100\%$ of the replicates return the count of that run (Sec.~\ref{sec:baselines}). Resampled consensus and stability assessment are established \cite{monti2003consensus,fred2005evidence,dudoit2003bagging,hennig2007clusterwise}, with known caveats as selection criteria \cite{vonluxburg2010stability}. Therefore, we use Monte Carlo resampling to certify the number of clusters that the occupancy produces; the resampling never selects it.

\begin{table}[!ht]
\caption{Parameters of the clustering algorithm and the values used in this work (seed $0$ for the subsample draws).}\label{tab:params}
\begin{ruledtabular}
\begin{tabular}{llll}
symbol & meaning & default & this work \\ \hline
$k$ & difference order & bounds of Sec.~\ref{sec:weights} & $1$ \\
$\varepsilon$ & filter resolution on $w$ & plateau of the count (Sec.~\ref{sec:baselines}) & $0.1$ \\
$\nmin$ & occupancy floor & $15$ & $15$ \\
$p$, $w_i$ & norm exponent and order weights, Eqs.~\eqref{def:Dsk} and \eqref{eq:weights} & $2$, $1$ & $2$, $1$ \\
$B$ & Monte Carlo replicates & $50$ & $50$ \\
$\phi$ & subsample fraction & $0.8$ & $0.8$ \\
\end{tabular}
\end{ruledtabular}
\end{table}

%Topic: The case study runs at one reference setting, and typing a new curve afterwards is cheap.
The analysis of the OGTT G-I cohort in this work uses the parameters in Table~\ref{tab:params}. Throughout, the weights are the data-driven scales of Eq.~\eqref{eq:weights} with $w_i\equiv1$ and $\alpha_0>0$, so $D_s^k$ is the metric of Proposition~\ref{prop:metric}; Appendix~\ref{ap:sensitivity} varies $\nmin$ and the weights.

\section{The OGTT cohort and its clinical criterion}\label{sec:cohort}\label{sec:casestudy}

%Topic: The cohort is 1283 paired five-point OGTTs from a single Chilean provider.
We analyze paired glycemia and insulinemia curves from five-point OGTTs ($N=1283$), collected in routine clinical practice at a reproductive-health clinic in Chile \cite{contreras2020novel,oliveranappa2022patient}. The cohort is $82.9\%$ female, with a median age of $33$~years (IQR $26$--$45$, range $18$--$81$), and the tests were run between $2008$ and $2015$. The grid is $t=0,30,60,90,120$~min, so $\vec G,\vec I\in\R^5$ at $\dt=30$~min, in \si{mg/dL} and \si{\micro U/mL}.

%Topic: The criterion is stricter than the standard reading, and its label is not a diagnosis.
The clinical normality criteria of Ref.~\cite{oliveranappa2022patient} split the cohort into $454$ clinically normal and $829$ abnormal subjects. A subject is normal if six conditions hold: fasting glucose $<100$~\si{mg/dL}; fasting insulin $<15$~\si{\micro U/mL}; glucose $\leq160$~\si{mg/dL} at every time point; $G(120)\leq140$~\si{mg/dL}; insulinemia
$\leq100$~\si{\micro U/mL} at every time point; $I(120)\leq60$~\si{\micro U/mL}. These criteria are stricter than the standard diagnostic reading by fasting glucose and the $2$-h value alone ($<100$ and $<140$~\si{mg/dL} normal, $140$--$199$ impaired tolerance, $\geq200$ diabetic) \cite{jagannathan2020oral}, as they also constrain the intermediate points and insulin. ``Abnormal'' here is a flag that the criteria associated with this test have been violated, not a diagnosis. We treat this criterion as a label to be explained rather than as a pre-processing step and cluster the cohort as a whole.

\section{Response types in glycemia and insulinemia and their implications for health}\label{sec:baselines}

\begin{figure}[tb]
    \centering
    \includegraphics[width=\linewidth]{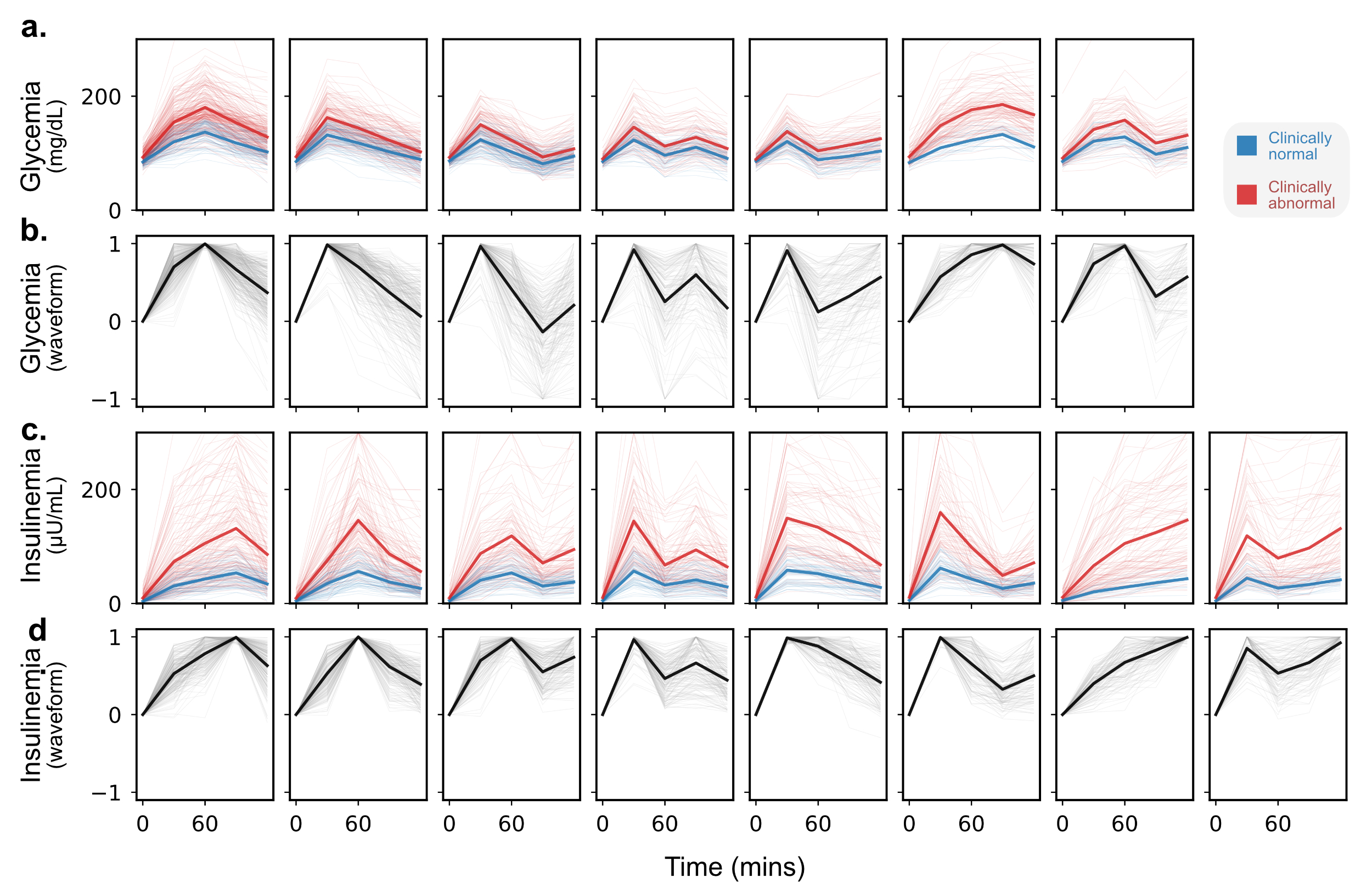}
    \caption{\textbf{Seven glucose and eight insulin waveform types cover the cohort, and within a type the two clinical classes share the shape and separate by level and excursion.} Rows a and c draw the curves of each type in \si{mg/dL} and \si{\micro U/mL}, colored by clinical class (blue normal, red abnormal) with the class means in bold; rows b and d draw the corresponding unit waveforms $w$, members in gray with the type mean in black. Columns are the types, largest first ($7$ glucose in a--b, $8$ insulin in c--d), at the reference setting ($k=1$, $\varepsilon=0.1$, seed $0$, $N=1283$). Reading down a column, the black waveform mean fixes the shape while the two class means in the curve row stay close to parallel; Table~\ref{tab:types} gives each type its trend pattern, size, normal/abnormal split, and health index.}
    \label{fig:atlas}
\end{figure}

%Topic: The reference setting returns 7 glucose and 8 insulin types that cover almost the whole cohort.
Running our algorithm with the parameters in Table~\ref{tab:params} on the cohort in Sec.~\ref{sec:cohort}, we found $7$ glucose and $8$ insulin response types (Fig.~\ref{fig:atlas}, Table~\ref{tab:types}). Coverage is $0.984$ for glucose and $0.999$ for insulin; the residual holds $21$ subjects for glucose and $1$ for insulin, with a residual frequency above $0.5$.

%Topic: The realizable patterns are what the encoding can express; the cohort populates only part of them, and that occupancy sets the number of types.
The $16$ realizable trend patterns at $k=1$ are the image set of the encoding once every neutral symbol is resolved, i.e., every response the method can express at this order. Of them, the cohort populates seven on glucose and eight on insulin above the occupancy floor, with uneven frequencies (Fig.~\ref{fig:atlas} and Table~\ref{tab:types}); the data thus show which realizable patterns this sample populates and how frequent they are. The occupancy is robust to the resolution and to the floor: the glucose count holds at $7$ across the threefold range $\varepsilon=0.05$--$0.15$ and the insulin count at $8$ across $\varepsilon=0.05$--$0.10$, with $98$--$100\%$ of the replicates returning the modal count on these plateaus. At $\varepsilon=0.1$, the glucose count also holds for $\nmin$ between $10$ and $30$, whereas the insulin count of $8$ requires $\nmin\leq15$ (Appendix~\ref{ap:sensitivity}). Altogether, the number of types is the occupancy of the realizable patterns, read at a difference order, a resolution, and an occupancy floor; on the plateaus, none of the three changes it.

%Topic: k=1 is the reference because the second-order occupancy stabilizes on glucose but not on insulin.
At $k=2$, the encoding also reads curvature trends, and the two variables respond differently. On glucose, the occupancy stabilizes again, at $11$ of the $54$ realizable patterns: the refinement splits three of the seven first-order types by late curvature, one of them into three branches of markedly different composition (the peak-at-$60$ type separates into branches whose clinically normal shares span $0.18$ to $0.41$). On insulin, no plateau emerges: the count briefly holds at $9$ over $\varepsilon=0.10$--$0.15$, but the replicate support there is erratic, with the modal count returning in $38$ to $98\%$ of the draws. For this reason, and per the window rule of Sec.~\ref{sec:weights}, we keep $k=1$ as the reference and report the curvature refinement without adopting it.

\begin{table}[!ht]
\caption{Waveform types at the reference setting, ordered by ascending health index $h$, i.e., the share of clinically normal members; $n$ is the type size, with its normal/abnormal decomposition in parentheses; the type numbers used in Figs.~\ref{fig:atlas}, \ref{fig:arms}, and~\ref{fig:conditional} follow descending $n$.}\label{tab:types}
\begin{ruledtabular}
\begin{tabular}{llrr}
type & reading & $n$ (N/A) & $h$ \\ \hline
\multicolumn{4}{l}{glucose ($g=7$)}\\
\texttt{+++-} & peak at 90 & 123 (21/102) & 0.17 \\
\texttt{++-{}-} & peak at 60 & 303 (78/225) & 0.26 \\
\texttt{++-+} & peak at 60, late rebound & 108 (38/70) & 0.35 \\
\texttt{+-{}-{}-} & peak at 30, falling & 250 (96/154) & 0.38 \\
\texttt{+-++} & early peak, rising again & 134 (53/81) & 0.40 \\
\texttt{+-+-} & biphasic & 143 (61/82) & 0.43 \\
\texttt{+-{}-+} & peak at 30, late rebound & 201 (96/105) & 0.48 \\
\hline
\multicolumn{4}{l}{insulin ($g=8$)}\\
\texttt{++++} & still rising at 120 & 123 (17/106) & 0.14 \\
\texttt{+-++} & early peak, rising again & 119 (36/83) & 0.30 \\
\texttt{+++-} & peak at 90 & 222 (72/150) & 0.32 \\
\texttt{++-{}-} & peak at 60 & 196 (75/121) & 0.38 \\
\texttt{++-+} & peak at 60, late rebound & 173 (69/104) & 0.40 \\
\texttt{+-{}-+} & peak at 30, late rebound & 135 (54/81) & 0.40 \\
\texttt{+-{}-{}-} & peak at 30, falling & 147 (60/87) & 0.41 \\
\texttt{+-+-} & biphasic & 167 (70/97) & 0.42 \\
\end{tabular}
\end{ruledtabular}
\end{table}

%Topic: The health index orders the types, from late peaks to late-rebound and biphasic responses.
After identifying the response types in Table~\ref{tab:types}, we explore their relationship with the threshold-based health criterion. We define the health index $h$ of a type as the share of its members who are clinically normal. The index varies by about a factor of $3$ across types, and its ordering is physiological: late peaks rank lowest (the glucose peak at $90$ and insulin still rising at $120$), while late-rebound and biphasic responses rank highest. No type is clinically pure, and types of similar $h$ may fail the criterion for different physiological reasons.

\begin{figure}[tb]
    \centering
    \includegraphics[width=130mm]{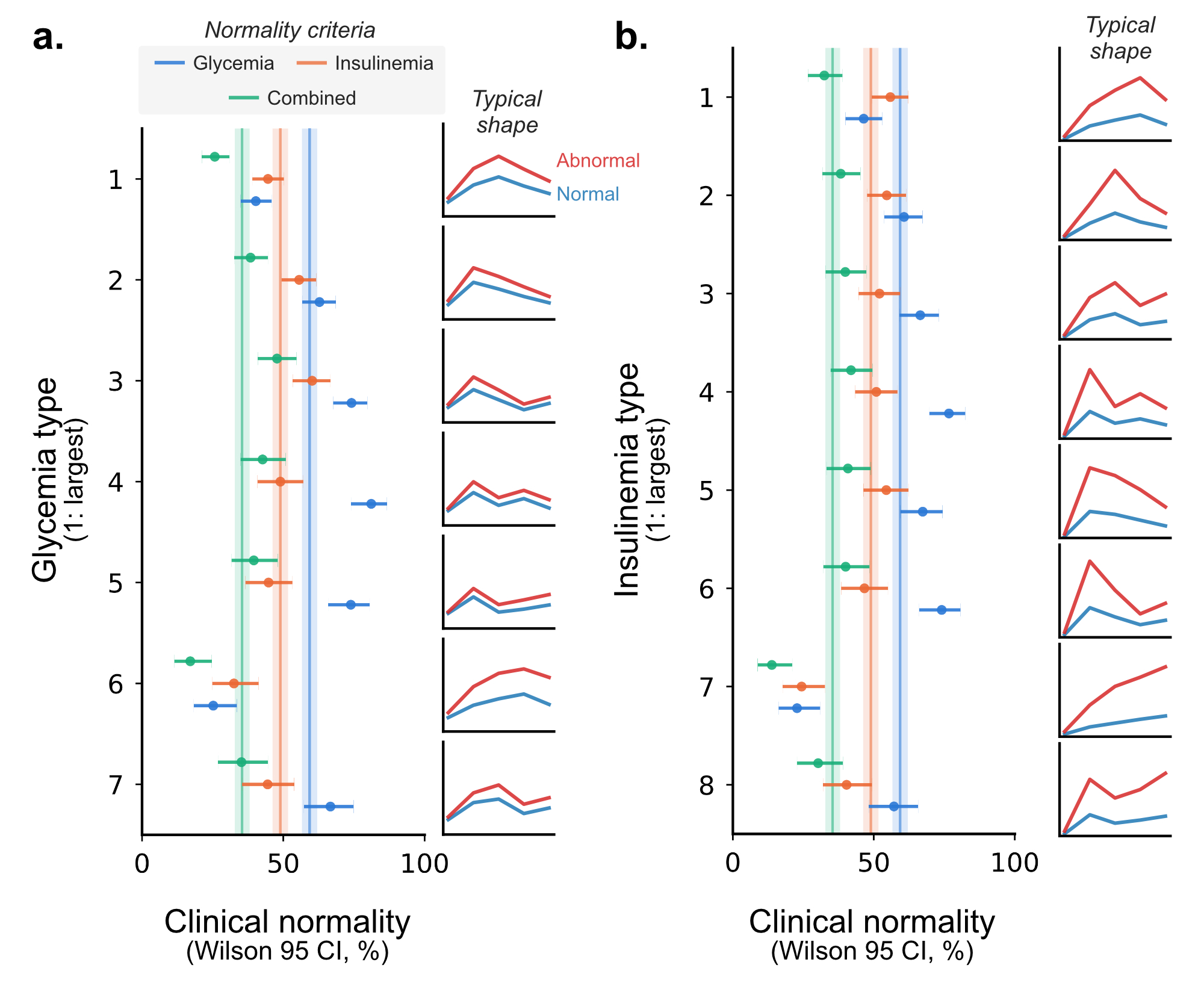}
    \caption{\textbf{The disagreement in health criteria is type-specific.} When the components of the health criteria are separated (blue, glycemia thresholds; orange, insulinemia thresholds; green, combined), the agreement between the glycemia and the insulinemia thresholds varies by type. Error bars denote the clinical normality and Wilson's $95\%$ confidence interval of each glycemia (\textbf{a}) and insulinemia (\textbf{b}) type, accompanied by their type's characteristic response profile (shape, next to each panel), sorted in descending size. }    
    \label{fig:arms}
\end{figure}

%Topic: Splitting the joint criterion into its glycemia and insulinemia thresholds locates where a type fails, and the split is itself type-specific.
Indeed, the health criterion combines thresholds for glycemia and insulinemia, and we can evaluate each set of thresholds separately on the same types to locate the source of failure (Fig.~\ref{fig:arms}). The glycemia thresholds are satisfied by $761$ of $1283$ subjects, the insulinemia thresholds by $628$ subjects, and both thresholds by $454$ subjects. The disagreement between the two contributions is itself type-specific. The glycemia thresholds pass a larger share in five of the seven glucose types and in six of the eight insulin types; in the remaining two per variable, the order reverses. The Wilson $95\%$ intervals separate cleanly in some types and overlap in others.

%Topic: can we recover the types if we do first the separation and then the clustering?
However, would we recover the same types if we split the cohort by clinical criteria and then rerun the clustering algorithm? We test this on glucose by rerunning the algorithm within each clinical class and comparing the class-restricted labels with the pooled labels. Normals alone recover $4$ of the $7$ types, missing the peak-at-$90$, the peak-at-$60$-with-late-rebound and the early-peak-rising-again types, which together pool $112$ normal members; abnormals recover all $7$. The normal subset is the smaller of the two ($454$ vs. $829$), so part of this gap is due to occupancy rather than vocabulary. Importantly, neither class produces a type absent from the pooled run. The class-restricted labels agree with the pooled labels at ARI $0.64$ for normals and $0.85$ for abnormals. The same waveform vocabulary thus spans both sides of the criterion; we interpret this as class-independence of the realizable trend patterns, not as validation.

\begin{figure}[tb]
    \centering
    \includegraphics[width=170mm]{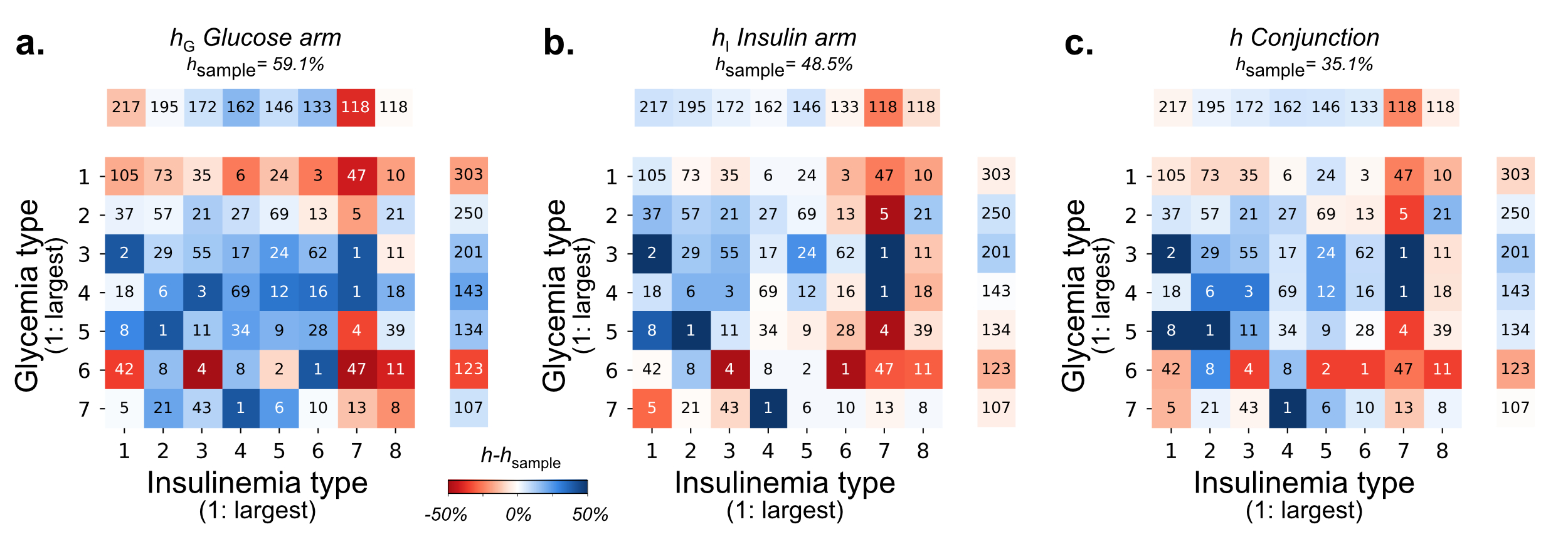}
    \caption{\textbf{Clinical normality is distributed across types, and no glycemia-insulinemia combination appears to carry health information beyond its two component types.} Cross-tabulation of glycemia against insulinemia types for the $1261$ of $1283$ subjects typed on both axes; each cell shows its membership count, and its color shows the difference in clinical normality relative to the general cohort, for glycemia (\textbf{a}), insulinemia (\textbf{b}), and combined criteria (\textbf{c}). The side strips give each type's marginal count and normal share. Intermediate shades dominate wherever counts are large; the deepest and palest cells contain eleven subjects or fewer and should be read as a sample-size effect.}
    \label{fig:conditional}
\end{figure}

%Topic: The two typings co-occur well beyond chance, but no specific combination moves health beyond what each axis predicts alone.
How strongly are the two typings related, and does a specific $G_i$-$I_j$ combination carry health information beyond what its two types provide separately? Both questions are based on the cross-tabulation of the typings, the $7\times8$ table of glucose type against insulin type, with each cell's membership count and normal share, for $N=1261$ of $1283$ subjects carrying a type label on both axes (Fig.~\ref{fig:conditional}). The $22$ excluded are markedly healthier than the rest ($54.5\%$ clinically normal against $35.0\%$), so they are not a representative subset. On the first question, the table is far from the product of its marginals: glucose and insulin types co-occur well beyond chance (bias-corrected Cram\'er's $V=0.350$, permutation $p=0.0002$, $B=5000$). On the second, health adds no structure that these data resolve. A cell-grouped additive logistic model, $\mathrm{logit}(p_{ij})=\alpha+\beta_G[i]+\beta_I[j]$, is not rejected against the cell-level saturated model: deviance $57.80$ with $42$ degrees of freedom, $p=0.116$ under a parametric bootstrap null ($B=2000$), which replaces the asymptotic $\chi^2$ that the sparse cells make unreliable. The margin is narrow; we read $p=0.116$ as a failure to detect an interaction at this sample size. We therefore detect no $G_i$-$I_j$ combination that pushes health beyond what its two types predict separately. The correlation itself is expected: each pair of curves is sampled from a single physiology, the person taking the test. What a reader might intuitively call ``an unhealthy combination'' is, in these data, this co-variation rather than a health interaction. 

\section{Benchmark against standard clustering methods}\label{sec:benchmark}
%Topic: We fix three rules before comparing methods: no clinical labels, no classifier metrics, internal indices only as diagnostics.
Here, we benchmark our method against six classical clustering methods and two time-series methods, $k$-Shape \cite{paparrizos2015kshape} and DTW $k$-means \cite{holder2024elastic}, at a matched number of groups, $g=7$ for glycemia and $g=8$ for insulinemia. Appendix~\ref{ap:benchmark} sweeps $g$ away from that choice. Three rules fix the comparison. First, every method is unsupervised, and the clinical criterion plays no role in the construction. Second, we report no classification performance: the types describe the excursion and do not predict a threshold-based clinical label. Third, we read the internal validity indices as diagnostics, never as evidence that one partition is better; each is defined by the distance the method itself optimizes. The benchmarked methods run on four representations: the raw curves, their $z$-scored and first-difference versions, and the waveform factorization of Sec.~\ref{sec:distance}; the two time-series methods run on the waveform. No method is therefore penalized by its representation alone (Figs.~\ref{fig:benchG} and \ref{fig:benchI} in the Supplemental Material show the groups of the six classical methods).

Table~\ref{tab:benchmark} reports two sets of quantities. The first is independent of $D_s^k(w)$: the reconstruction error, the stability of the partition, the variance in the mean level explained by the group assignment ($\eta^2$), and the share of the cohort in the largest group. The second is the trend mismatch, defined through the filtered patterns of Sec.~\ref{sec:filter}. Every curve is reconstructed from the mean waveform $\bar w_{c(i)}$ of the group $c(i)$ it belongs to, placed at its own resting level $x_{i0}$ and amplitude $A_i$,
\begin{equation}
  \mathrm{RMSE}=\sqrt{\frac{1}{4N}\sum_{i=1}^{N}\sum_{t=1}^{4}\bigl(x_{it}-\hat x_{it}\bigr)^{2}},
  \qquad
  \hat x_{i}=x_{i0}+A_i\,\bar w_{c(i)},
  \label{eq:rmse}
\end{equation}
with $N$ being the number of curves. The sum runs over the four informative points, since $\bar w_{c0}=0$ matches the fasting value exactly for every method. Stability is the subsample ARI ($B=50$, $\phi=0.8$, for every method), and $\eta^2$ is the between-group over the total sum of squares of the mean level.

All methods produce groups that reduce the reconstruction error, though to varying degrees. The row without grouping sets the scale for reading their spread: a single mean waveform, with no types at all, already reconstructs the cohort at $23.4$~\si{mg/dL} and $37.1$~\si{\micro U/mL}. The distances between methods are therefore a fraction of what the factorization alone leaves, and for glycemia, they remain below the test's reproducibility (Sec.~\ref{sec:filter}). For glycemia, our method is the most stable among the methods compared and has the lowest $\eta^2$ among those that preserve the waveform amplitude, marginally below that of the GMM; $k$-Shape standardizes each waveform again and reaches $0.208$. Its reconstruction error falls within the range of the benchmarked methods. The low $\eta^2$ confirms that our types respond to the shape of the excursion rather than to its mean level, i.e., the division of labor intended in Sec.~\ref{sec:distance}. For insulinemia, the most stable method is the GMM, which has a larger reconstruction error than ours and places a third of the cohort in a single component. The two time-series methods are among the least stable for both variables, and $k$-Shape places $42\%$ of the insulin cohort into a single group. On five points, there is little to align; the shift and the warping that their invariances remove are exactly what the protocol grid holds fixed. 

Our types are trend-preserving by construction at the working resolution, over the curves they assign. Every benchmarked method, in contrast, groups curves whose trends contradict the consensus within their own group; at $\varepsilon=0.1$, this affects $17.2$--$38.8\%$ of the cohort for glycemia and $4.5$--$29.7\%$for on insulinemia, including the two time-series methods. Part of that mismatch lies with the assignment rather than with the partition. We therefore apply our assignment rule to every method, moving each curve to the nearest group whose consensus it does not contradict. The shares fall to $6.1$--$22.4\%$ and $0.1$--$12.5\%$, while ours rise from zero to $1.6$ and $0.1\%$. Requiring an exact match of signs (i.e., imposing $\varepsilon=0$) roughly doubles the glycemia shares and multiplies the insulinemia shares by factors between two and ten; our own reading moves from $0$ to $15.3$ and $22.5\%$. The zero is therefore a statement at the working resolution, and the resolution enters insulinemia more strongly than glycemia. Mixing contradictory trends is problematic in itself, since such groups combine individuals whose responses returned to the resting state in different ways, potentially reflecting distinct underlying physiological mechanisms.

\begin{table}[!ht]
\caption{Benchmark at matched count ($g=7$ glucose, $g=8$ insulin; reference setting). Every method assigns the entire cohort except ours, leaving the $21$ glucose and $1$ insulin curves unassigned, a pattern compatible with none of the selected patterns.}\label{tab:benchmark}
\begin{ruledtabular}
\begin{tabular}{lrrrr|rrr}
 & \multicolumn{4}{c|}{independent of $D_s^k(w)$} & \multicolumn{3}{c}{trend mismatch (\%)} \\
\cline{2-5}\cline{6-8}
method & RMSE & stab. & $\eta^2$ & largest & \multicolumn{1}{c}{$\varepsilon\!=\!0.1$} & \multicolumn{1}{c}{reassigned} & \multicolumn{1}{c}{$\varepsilon\!=\!0$} \\ \hline
\multicolumn{8}{l}{glucose ($g=7$)}\\
\emph{no grouping} ($g=1$) & 23.4 & --- & --- & 100\% & --- & --- & --- \\
ours & 15.9 & 0.818 & 0.262 & 23.6\% & 0.0 & 1.6 & 15.3 \\
\hline
$k$-means, raw & 16.6 & 0.681 & 0.914 & 27.7\% & 38.8 & 22.4 & 63.3 \\
$k$-means, $z$-scored & 12.9 & 0.626 & 0.317 & 21.0\% & 17.2 & 14.3 & 41.5 \\
$k$-means, differences & 17.4 & 0.579 & 0.440 & 23.1\% & 21.7 & 6.9 & 47.9 \\
$k$-means, waveform & 12.5 & 0.654 & 0.358 & 27.4\% & 28.4 & 10.4 & 52.6 \\
GMM, waveform & 15.3 & 0.774 & 0.267 & 32.6\% & 20.7 & 20.3 & 40.9 \\
Ward, waveform & 13.2 & 0.446 & 0.347 & 32.6\% & 26.3 & 20.3 & 56.2 \\
$k$-Shape, waveform & 17.5 & 0.482 & 0.208 & 30.1\% & 24.0 & 6.1 & 56.8 \\
DTW $k$-means, waveform & 14.8 & 0.465 & 0.321 & 27.0\% & 38.2 & 13.7 & 64.8 \\
\hline
\multicolumn{8}{l}{insulin ($g=8$)}\\
\emph{no grouping} ($g=1$) & 37.1 & --- & --- & 100\% & --- & --- & --- \\
ours & 24.1 & 0.738 & 0.025 & 17.3\% & 0.0 & 0.1 & 22.5 \\
\hline
$k$-means, raw & 29.2 & 0.797 & 0.919 & 36.3\% & 15.8 & 12.5 & 75.4 \\
$k$-means, $z$-scored & 19.7 & 0.732 & 0.029 & 17.2\% & 12.5 & 1.9 & 53.2 \\
$k$-means, differences & 28.4 & 0.796 & 0.542 & 49.2\% & 11.4 & 2.8 & 66.7 \\
$k$-means, waveform & 19.0 & 0.616 & 0.023 & 19.2\% & 14.5 & 3.3 & 55.6 \\
GMM, waveform & 26.2 & 0.913 & 0.014 & 32.7\% & 4.5 & 3.0 & 44.8 \\
Ward, waveform & 20.6 & 0.362 & 0.019 & 19.7\% & 13.3 & 3.5 & 53.5 \\
$k$-Shape, waveform & 25.8 & 0.401 & 0.014 & 41.7\% & 9.4 & 0.1 & 62.9 \\
DTW $k$-means, waveform & 24.3 & 0.518 & 0.015 & 19.6\% & 29.7 & 9.9 & 67.4 \\
\end{tabular}
\end{ruledtabular}
\end{table}

%Topic: The residual grows along the curve for the trend-responsive partitions and stays flat for the rest, which is the trade the types make.
The different methods do not weigh the temporal points equally, and the accumulation of reconstruction error along the curve distinguishes what each partition groups on (Fig.~\ref{fig:residual_profile} in the Supplemental Material). The reason is structural: a trend pattern fixes the sign of each segment and leaves its size free, so the members of a type agree on the way they return toward the resting level and differ in the amplitude of that return. Accordingly, on glycemia, the residual grows steadily along the curve for ours and for $k$-means on first differences, and for $k$-Shape, the partitions respond to the direction of change; the methods that group the excursion as a whole keep it nearly flat. The types thus buy agreement on the direction of every segment at the price of the return's amplitude.

%Topic: The GMM's performance metrics look competitive but its composition is poor; STEM and DIB-C are not directly comparable entries.
The GMM on waveforms requires a separate reading, since its performance metrics appear competitive, whereas its composition does not. Three observations support this. First, a third of the cohort concentrates in a single component. Second, a fifth of the glycemia cohort contradicts the trends of its own component. Third, its own BIC, when unconstrained, selects $5$ components instead of $7$. STEM and DIB-C, the closest precursors of Sec.~\ref{sec:filter}, are absent from Table~\ref{tab:benchmark} because their filters are defined on replicates or on a fixed transition level. With its own transition level and our minimum occupancy $\nmin=15$, DIB-C returns six glucose and eight insulin groups, with half the glucose cohort in its largest group; it is therefore not directly comparable at matched count under the waveform normalization. Finally, the matched count is itself a choice; Appendix~\ref{ap:benchmark} sweeps $g$ away from it and shows that no selection criterion available to the benchmarked methods points to $g=7$ or $g=8$.

\section{Discussion and outlook}\label{sec:conclusions}

%Topic: Undersampled observation of a complex system motivates the method; the case study shows what it finds.
In many situations, the resolution at which a complex system can be observed is not sufficient to infer its internal state, for economic, ethical, or practical reasons. Routine clinical tests are among them: in the OGTT, a fasting individual receives a controlled glucose load, and the system's response is recorded at five time points. Curves this short cannot always identify the dynamics behind them, yet they retain the nature of the response. Grouping them by that nature requires a clustering method that keeps the local trends of the response, which we propose here. Our method treats each curve as an undersampled excursion after the perturbation and retains its trends, i.e., the signs of its finite differences above the resolution, as the observables; it then compares shapes using a discrete Sobolev-type distance on the factorized waveform. In the case study, it discovered seven glucose and eight insulin response types, each a distinct progression of the response over the protocol window.

%Topic: The trend patterns turn the count into an occupancy, and the data populate only a few of them.
Filtering trend patterns is what made these observations possible. On a coarse grid, the magnitudes of the finite differences carry $O(\dt)$ uncertainty, whereas their signs are exact whenever the derivative keeps one sign across the window (Corollary~\ref{cor:sign}b); under noise, only the signs above the resolution are kept. The candidate types are therefore the realizable trend patterns and not an open-ended cluster set. Depending on the order of differences considered, there are $2^{n-1}=16$ fully specified patterns at $k=1$ and $2\cdot3^{n-2}=54$ at $k=2$ (Proposition~\ref{prop:count}). The cohort populates only a fraction of them, showing which patterns this physiology realizes in practice. Every reported type rises from fasting, which restricts the effective candidates to the eight patterns with a leading $+$; glucose populates seven of these and insulin all eight, and these occupancies are where the counts of Table~\ref{tab:types} come from. Within every type, the clinical criterion (threshold-based) acts as an amplitude cut inside a waveform rather than as a partition between waveforms.

%Topic: The distance resolves the curves the trends leave ambiguous, and classical methods are not constrained to preserve any of this.
The distance is what turns compatible patterns into a partition. A difference smaller than the resolution receives the neutral symbol $0$; the curve is then assigned by $D_s^k$ to the nearest medoid among the types its own trends admit. Curves whose trends are incompatible with every chosen type remain in the residual group ($21$ subjects on glucose, $1$ on insulin) rather than being forced into a group that misrepresents them. Classical methods are not constrained to preserve this structure, since a value-based distance reads level and amplitude. At a matched group count, every benchmarked method builds at least one group that mixes rising and falling segments, whereas ours builds none, by construction (Table~\ref{tab:benchmark}). From a complexity point of view, the method is efficient: the candidate set is finite and small, the cover works on at most $2\cdot3^{n-2}$ patterns at $k=2$, and the distance acts on the differences up to order $k$, so the full reference analysis, Monte Carlo certificate included, runs in minutes on a workstation.

%Topic: What limits the method, and what limits the data.
However, our results have limitations, some stemming from the method and others from the data. The method presupposes a controlled excursion from a resting state observed on a fixed window, and it is restricted to short series. On synthetic excursions over the same window, the populated patterns grow far slower than the realizable ones; what limits $n$ is the resolution, since with more points the change of the curve over one step falls below the noise, and the coverage collapses between $n=9$ and $n=13$ (Appendix~\ref{ap:sensitivity}). Where curves are long enough for alignment, value-based methods remain the right tools (Sec.~\ref{sec:benchmark}). Within that scope, the types are an occupancy, so a different sample may populate a different subset of the realizable patterns. The count is validated only where the resolution curve exhibits a plateau; insulin at $k=2$ shows none, so its curvature refinement is reported but never used as the reference. The types are also more reproducible than the individual labels. Under noise of the size of the repeat-test variability, the shapes persist, and the count moves by at most one, while more than half of the glucose curves, more often those of small excursion, change type (Appendix~\ref{ap:sensitivity}). The data impose limits of their own. The cohort is cross-sectional, comes from a single Chilean center, is mostly female, and is $64.6\%$ abnormal by referral, so no health index is a measured risk, and the observed proportions should not be interpreted as (population-based) epidemiological parameters, e.g., prevalence. Furthermore, the clinical criterion we grade against is stricter than typical fasting and 2h values \cite{jagannathan2020oral}, and a type's health index reflects in part the composition of the sample. The types reported here are, in short, those findable in such a sample. A larger and more diverse cohort, ideally with follow-up, would support more definitive statements about glycemic control and its development over time, how the shape of the response relates to health, and whether the encoding transfers to it without recalibration.

%Topic: Where else the framework applies, and what it carries there.
Finally, we would like to highlight that our method is not limited to applying to the OGTT; the same reasoning applies to any undersampled response of a system to a standardized perturbation. This scope includes other dynamic tests in clinical settings, as well as short segments of longer series, such as the growth phases of seasonal epidemics. On the theoretical side, the number of realizable patterns a cohort populates as the resolution varies invites a statistical mechanics of trend patterns, with occupation numbers plotted against $\varepsilon$ in the spirit of ordinal-pattern statistics \cite{bandt2002permutation,zanin2012permutation}. Altogether, the method extracts dynamic features from the curves and discovers which of the possible trend patterns are realized in the data. These patterns can facilitate parameter identification, e.g., by providing candidates for local regularization, and may shed light on the dynamics of other systems.

\section*{Ethics statement}
We used the ecological data described in \cite{oliveranappa2022patient}. The studies involving human participants were reviewed and approved by the Ethical Committee of Mutual de Seguridad, Santiago, Chile, under resolution case \#78. 

\section*{Data and code availability}
%Topic: The data come on request and the implementation will be public on publication.
The datasets analyzed are available from the corresponding authors upon reasonable request. The reference implementation of the method, together with all analysis and benchmark scripts, is available on GitHub \url{https://github.com/dmonarchc/stsc-ogtt}.

\begin{acknowledgments}
DM-O acknowledges funding from FONDECYT Iniciación 11250295. Claude Code was used as a coding assistant, and Claude Fable was used for grammar and literature checks. All AI-assisted code was manually reviewed, tested, and validated by the authors. The authors have thoroughly reviewed the final document and take full responsibility for its content.
\end{acknowledgments}

%TC:ignore

\appendix

\section{Proof of Lemma~\ref{lem:mvt}}\label{ap:proofs}

%Topic: The proof of the lemma combines two classical facts.
The proof combines two classical facts \cite{deboor2005divided}.

%Topic: Forward differences are scaled divided differences.
\emph{Step 1: forward differences are scaled divided differences.} Writing $E$ for the shift operator, $(Ef)(t)=f(t+\dt)$, the first difference is $\Delta = E-I$ and the $m$-th is its power, $\Delta^m=(E-I)^m$; powers factor, so $\Delta^m=\Delta\circ\Delta^{m-1}$. With $f[t_i]=f(t_i)$ and $f[t_i,\dots,t_{i+m}]=\big(f[t_{i+1},\dots,t_{i+m}]-f[t_i,\dots,t_{i+m-1}]\big)/(t_{i+m}-t_i)$, induction on $m$ gives, on a uniform grid,
\begin{equation}\label{eq:fd-dd}
\Delta^m[f](t_i) = m!\,(\dt)^m\, f[t_i,\dots,t_{i+m}].
\end{equation}
For $m=1$ both sides equal $f(t_{i+1})-f(t_i)$; assuming Eq.~\eqref{eq:fd-dd} for $m-1$,
\begin{align*}
\Delta^m[f](t_i) &= \Delta^{m-1}[f](t_{i+1})-\Delta^{m-1}[f](t_i)\\
&= (m-1)!\,(\dt)^{m-1}\big(f[t_{i+1},\dots,t_{i+m}]-f[t_i,\dots,t_{i+m-1}]\big)\\
&= (m-1)!\,(\dt)^{m-1}\,(t_{i+m}-t_i)\,f[t_i,\dots,t_{i+m}]
= m!\,(\dt)^{m}\,f[t_i,\dots,t_{i+m}].
\end{align*}

%Topic: Divided differences obey a mean-value theorem.
\emph{Step 2: mean-value theorem for divided differences.} For $f\in\mathcal{C}^m([t_i,t_{i+m}])$ there exists $\xi\in(t_i,t_{i+m})$ with
\begin{equation}\label{eq:dd-mvt}
f[t_i,\dots,t_{i+m}]=\frac{f^{(m)}(\xi)}{m!}.
\end{equation}
Let $P$ be the interpolating polynomial of degree $\leq m$ through $(t_j,f(t_j))_{j=i}^{i+m}$; its leading coefficient, read off the Newton form of $P$, is $f[t_i,\dots,t_{i+m}]$. The error $e:=f-P$ vanishes at the $m+1$ \emph{distinct} nodes $t_i<\dots<t_{i+m}$. By Rolle's theorem, $e'$ has at least $m$ distinct zeros, one strictly inside each of the $m$ gaps, hence all in the open interval $(t_i,t_{i+m})$. Inductively, for $1\leq r\leq m$, $e^{(r)}$ has at least $m+1-r$ distinct zeros in $(t_i,t_{i+m})$, the strict interlacing preserving distinctness at every stage; $r=m$ yields $\xi\in(t_i,t_{i+m})$ with $e^{(m)}(\xi)=0$. Since $P^{(m)}\equiv m!\,f[t_i,\dots,t_{i+m}]$, evaluating $e^{(m)}(\xi)=0$ gives Eq.~\eqref{eq:dd-mvt}.

%Topic: The kernel of the difference operator is the grid restriction of the low-degree polynomials.
\emph{The kernel of $\Delta^j$.} Each forward-difference map $\Delta^1:\R^{\ell}\to\R^{\ell-1}$ is surjective, cumulative summation providing a preimage, so $\Delta^j:\R^n\to\R^{n-j}$ is surjective and $\dim\ker\Delta^j=j$ by rank--nullity. The restriction map $\mathcal{P}_{j-1}\to\R^n$ is injective, a nonzero polynomial of degree $\leq j-1$ having at most $j-1$ roots on a grid of $n\geq j+1$ points, and its image lies in $\ker\Delta^j$ by Lemma~\ref{lem:mvt}; equality follows by dimension. Finally $\ker\Delta^j\subseteq\ker\Delta^i$ for $i\geq j$, because $\Delta^i=\Delta^{i-j}\circ\Delta^j$, so $D_s^k(\vec x,\vec y)=0$ exactly when $\Delta^j[\vec x-\vec y]=0$.

%Topic: The two steps together give the lemma, its corollary, and the magnitude bound.
Combining Eqs.~\eqref{eq:fd-dd} and \eqref{eq:dd-mvt} yields Lemma~\ref{lem:mvt}; Corollary~\ref{cor:sign} follows since $(\dt)^m>0$. For Remark~\ref{rem:magnitude}, write $f^{(m)}(\xi)-f^{(m)}(t_i)=f^{(m+1)}(\eta)(\xi-t_i)$ for some $\eta$ between $t_i$ and $\xi$, and bound $\abs{\xi-t_i}<m\,\dt$.

\section{Uneven grids}\label{ap:uneven}

%Topic: Scaled divided differences replace forward differences on an uneven grid.
The framework applies to both uniform and non-uniform grids. For strictly increasing nodes $t_0 < \dots < t_{n-1}$, we define the scaled divided difference
\begin{equation}\label{def:dd}
\mathfrak d^m_j[\vec x] \;:=\; m!\,x[t_j,\dots,t_{j+m}],\qquad j\leq n-1-m,
\end{equation}
where $x[\cdot]$ denotes the standard divided difference. On a uniform grid, this is exactly $\Delta^m[\vec x](t_j)/(\Delta t)^m$.

%Topic: The mean-value structure survives on uneven grids.
The mean-value result from Sec.~\ref{sec:distance} remains valid, as the proof of Lemma~\ref{lem:mvt} does not require uniform spacing. For $f \in \mathcal{C}^{m}([t_j, t_{j+m}])$, there exists $\xi \in (t_j, t_{j+m})$ such that $\mathfrak d^m_j[f] = f^{(m)}(\xi)$, so Corollary~\ref{cor:sign} remains unchanged. Remark~\ref{rem:magnitude} now reads $|\mathfrak d^m_j[f] - f^{(m)}(t_j)| \leq (t_{j+m}-t_j) \sup|f^{(m+1)}|$, with the window length $t_{j+m}-t_j$ replacing $m\,\Delta t$.

%Topic: The distance, the weights, and the order bound all follow from the divided differences.
All subsequent definitions and results follow from Eq.~\eqref{def:dd}. In the distance formula, $\Delta^i[\vec x]/(\Delta t)^i$ is replaced by $(\mathfrak d^i_r[\vec x])_r$, and Proposition~\ref{prop:metric} holds by the same argument, as $\mathfrak d^i$ is linear and its kernel remains the set of grid restrictions of $\mathcal P_{j-1}$. The weights are computed by pooling $|\mathfrak d^i_j|$. Equation~\eqref{Shannon} uses $\max_j(t_{j+1}-t_j)$ in place of $\Delta t$. The reference implementation allows for an explicit node vector and switches automatically between uniform and non-uniform grids.

%Topic: An uneven grid costs the sign filter its uniform order factor and worsens crowding locally.
Two practical differences should be noted. First, while the threshold remains $\varepsilon_m$ on $w$, the variance of $\mathfrak d^m_j$ is position-dependent on an uneven grid, so the usual order factor $\sqrt{\binom{2m}{m}/2}$ becomes specific to each position. Second, this variance increases further where nodes are closely spaced, as divided differences involve small node gaps; the local variance formula is given in Sec.~\ref{sec:weights}, with reciprocal products of node separations replacing $(\Delta t)^{-2m}$.

\section{How many trend patterns are realizable}\label{ap:count}

%Topic: Realizability is a chain of sign constraints on the first differences, checkable by forward propagation.
Not every string of symbols is the trend pattern of an actual curve: Sign compatibility constraints need to be imposed for higher difference orders, because not all patterns are realizable. For example, a rising step followed by a falling one therefore forces a negative curvature between them, so that the string can only be a $-$. How many of the $3^K$ formal patterns are realizable, i.e., produced by at least one $\vec x\in\R^n$, is a property of the protocol and the resolution threshold, and at $k=2$ it can be counted exactly. To do so, write $v_j:=\Delta^1_j[\vec x]$ for the first differences. A $k=2$ pattern prescribes the sign of every $v_j$ (its slope symbols) and of every $v_{j+1}-v_j$ (its curvature symbols), so it is realizable exactly when these prescriptions, read as inequalities on the real numbers $v_0,\dots,v_{n-2}$, admit a solution; the level $x_0$ never enters, since any difference vector is realized by cumulative summation from any starting value. Realizability can thus be checked by propagating the constraints forward: the first slope symbol fixes the admissible set for $v_0$, each subsequent curvature-slope pair maps the admissible set for $v_j$ onto one for $v_{j+1}$, and the pattern is realizable when no step returns the empty set.

%Topic: At $\varepsilon=0$ the propagation only remembers the current sign, and counting patterns reduces to a $3\times3$ transfer matrix.
At $\varepsilon=0$, the admissible set for each $v_j$ is always one of the three sign sets $(0,\infty)$, $\{0\}$, $(-\infty,0)$ in full, so the only memory the chain keeps is the current sign of $v_j$. Since that sign takes three values, the bookkeeping is a $3\times3$ matrix whose rows index the sign of $v_j$, whose columns index the sign of $v_{j+1}$, and whose entries count the curvature symbols consistent with that move,
\begin{equation}\label{eq:transfer}
M=\bordermatrix{ & + & \;0\; & - \cr
+ & 3 & 1 & 1 \cr
0 & 1 & 1 & 1 \cr
- & 1 & 1 & 3 }.
\end{equation}
For example, $M_{11}=3$, because if $\sgn(v_j)=\sgn(v_{j+1})=+$, then their difference can have any sign, whereas $M_{12}=1$, because if $\sgn(v_j)=+$ and $\sgn(v_{j+1})=0$, then $\sgn(v_{j+1}-v_j)$ can only be $-$. A realizable pattern is then a starting sign followed by $n-2$ steps, each choosing the next sign and one of the curvatures its entry of $M$ admits; because every admissible set is a full sign set, choices at different steps are independent, and summing over all starting signs and all walks gives the matrix power of Proposition~\ref{prop:count}. The largest eigenvalue of $M$ is, in turn, the factor by which each added measurement multiplies the count. The patterns excluded by this count are forbidden by the encoding and not by the dynamics, so their absence carries none of the information of forbidden patterns in the sense of Ref.~\cite{amigo2008forbidden}; what the realizable set fixes is the resolution available to the analysis.

\begin{prop}[Realizable trend patterns]\label{prop:count}
Let $A(n)$ be the number of patterns $\pi\in\{-1,0,+1\}^{K}$ realized by at least one $\vec x\in\R^n$ under Eq.~\eqref{Filtro} with $k=2$ and $\varepsilon=0$. Then
\begin{equation}\label{eq:count}
A(n)=\bm 1^{\!\top}M^{\,n-2}\bm 1,
\end{equation}
with $M$ the transfer matrix of Eq.~\eqref{eq:transfer}. The characteristic polynomial of $M$ factors as $(x-2)(x^2-5x+2)$, so its leading eigenvalue is $\lambda=(5+\sqrt{17})/2\approx4.562$. Thus, $A(n)\sim C\lambda^n$ against $3^{K}=9^{n}/27$ formally (Eq.~\ref{eq:pattern}). Curves in general position, having no exact ties, realize exactly $2\cdot3^{\,n-2}$ patterns, all of them with symbols in $\{-1,+1\}$.
\end{prop}

\begin{proof}
At $\varepsilon=0$ every threshold vanishes, so each symbol confines its quantity to one of the three sets $(0,\infty)$, $\{0\}$, $(-\infty,0)$. Checking the $27$ sign triples shows that the propagated set is always one of these three in full, never a fragment of one, so the choices at successive steps are independent and the admissible transitions are exactly counted by the entries of $M$ in Eq.~\eqref{eq:transfer}. Each pattern is a starting sign followed by $n-2$ steps, and it may start and end at any sign; summing over the walks is the matrix power, and summing over the two free ends contracts it with $\bm 1$ on both sides, which gives Eq.~\eqref{eq:count}. Being symmetric, $M$ is diagonalizable and admits an orthonormal eigenbasis $\{u_i\}$: writing $M=PDP^{\!\top}$ with the $u_i$ as columns of $P$, the power telescopes to $M^{\,n-2}=P\,D^{\,n-2}P^{\!\top}$, and Eq.~\eqref{eq:count} becomes $A(n)=\sum_i \big(u_i^{\!\top}\bm 1\big)^2\lambda_i^{\,n-2}$. The characteristic polynomial of $M$ factors as $(x-2)(x^2-5x+2)$. The eigenvector of $\lambda=2$ is proportional to $(1,0,-1)^{\!\top}$, orthogonal to $\bm 1$, so its weight vanishes; $A(n)$ therefore follows the two roots of $x^2-5x+2$, and asymptotically the leading one, $A(n)\sim C\lambda^{n}$ with $\lambda=(5+\sqrt{17})/2$. In general position all $v_j\neq0$ and all $v_{j+1}\neq v_j$, so the state $0$ and the curvature symbol $0$ disappear, and the bookkeeping matrix reduces to
$M_{\mathrm{gp}}=\left(\begin{smallmatrix}2&1\\1&2\end{smallmatrix}\right)$
on the signs $+,-$. The same reasoning applies: the eigenvector of $\lambda=1$ is proportional to $(1,-1)^{\!\top}$, orthogonal to $\bm 1$, so only the leading eigenvalue $\lambda=3$ contributes, with weight $\big((1,1)\tfrac{1}{\sqrt2}\cdot\bm 1\big)^2=2$, and the count is exactly $A_{\mathrm{gp}}(n)=\bm 1^{\!\top}M_{\mathrm{gp}}^{\,n-2}\bm 1=2\cdot3^{\,n-2}$.
\end{proof}

\section{Benchmark away from the matched count}\label{ap:benchmark}
%Topic: The appendix defines the pseudo-$F$ on a distance matrix; the other diagnostics are standard.
This appendix extends the benchmark of Sec.~\ref{sec:benchmark} beyond the matched group count. Of the internal diagnostics used here, only the pseudo-$F$ requires a definition for a general distance matrix. Writing $d_{ij}=D_s^k(w_i,w_j)$ and $C_1,\dots,C_g$ for the groups of a partition of $N$ curves, the total and within-group dispersions follow from the pairwise distances alone,
\begin{equation}
    \mathrm{SS}_{\mathrm{T}}=\frac{1}{N}\sum_{i<j} d_{ij}^{2},
    \qquad
    \mathrm{SS}_{\mathrm{W}}=\sum_{r=1}^{g}\frac{1}{n_r}\sum_{\substack{i<j\\ i,j\in C_r}} d_{ij}^{2},
\end{equation}
with $n_r=|C_r|$, and the pseudo-$F$ is the between-to-within ratio per degree of freedom,
\begin{equation}\label{eq:pseudoF}
    F=\frac{(\mathrm{SS}_{\mathrm{T}}-\mathrm{SS}_{\mathrm{W}})/(g-1)}{\mathrm{SS}_{\mathrm{W}}/(N-g)},
\end{equation}
the distance-based generalization of the Calinski--Harabasz index \cite{calinski1974dendrite,anderson2001new}. Larger values indicate compact, mutually separated groups under $D_s^k(w)$; every method's partition is evaluated with the same distance matrix.

%Topic: Away from the matched count, reconstruction rewards ever more groups while every selection criterion favors very few.
The matched count is a choice, and the comparison should survive it. Figure~\ref{fig:groupsweep} in the Supplemental Material sweeps $g=2,\dots,20$ for the six classical methods against two sets of diagnostics. The first is independent of $D_s^k(w)$: reconstruction rewards finer partitions, with RMSE falling monotonically for every benchmarked method, while stability decays with $g$, and five of the six native selection criteria choose $g=2$ on both variables; the sixth, the GMM's BIC, stops at $5$ and $4$. The second set evaluates the same partitions under $D_s^k(w)$: silhouette, pseudo-$F$, and the gap statistic peak at $g=2$--$4$ and decline beyond, whereas BIC-penalized reconstruction, although built to penalize added groups, keeps improving toward $g\approx20$; over this range, the reconstruction gain outpaces the penalty. No criterion in either set would have selected anything near $g=7$ or $g=8$. Figure~\ref{fig:residual_profile} complements the sweep at the matched count, decomposing the reconstruction error of Table~\ref{tab:benchmark} along the protocol grid.

%Topic: Our count comes from occupancy, and it still lies in the upper half of the spread.
Ours is fixed by the occupancy argument of Sec.~\ref{sec:baselines} and by none of these criteria; at $g=7$ and $g=8$ it lies beyond the optimum of the validity diagnostics and below the counts that reconstruction alone would favor. Even so, at the matched count our partition lies in the upper half of the six-method spread on all three validity diagnostics under $D_s^k(w)$, on both variables: the silhouette reads $0.138$ against a mean of $0.086$ over the benchmarked methods on glucose and $0.178$ against $0.056$ on insulin; pseudo-$F$, $253.5$ against $199.6$ and $329.2$ against $223.0$; the gap statistic, $0.866$ against $0.750$ and $0.414$ against $0.275$. Its BIC-penalized reconstruction lies within range but on the costlier side of the mean ($9053$ vs $8838$ for glucose, $10174$ vs $10106$ for insulin), consistent with that diagnostic's drift toward large $g$. Altogether, the group count is not a free parameter tuned to appear favorable against these benchmarked methods; wherever their own model selection would send them, it is not toward ours.

\section{Sensitivity and robustness analysis}\label{ap:sensitivity}

%Topic: The count survives the occupancy floor on the plateau of the resolution, and the weights of the distance only move the ambiguous curves.

\begin{figure}[!ht]
  \centering
  \includegraphics[width=\linewidth]{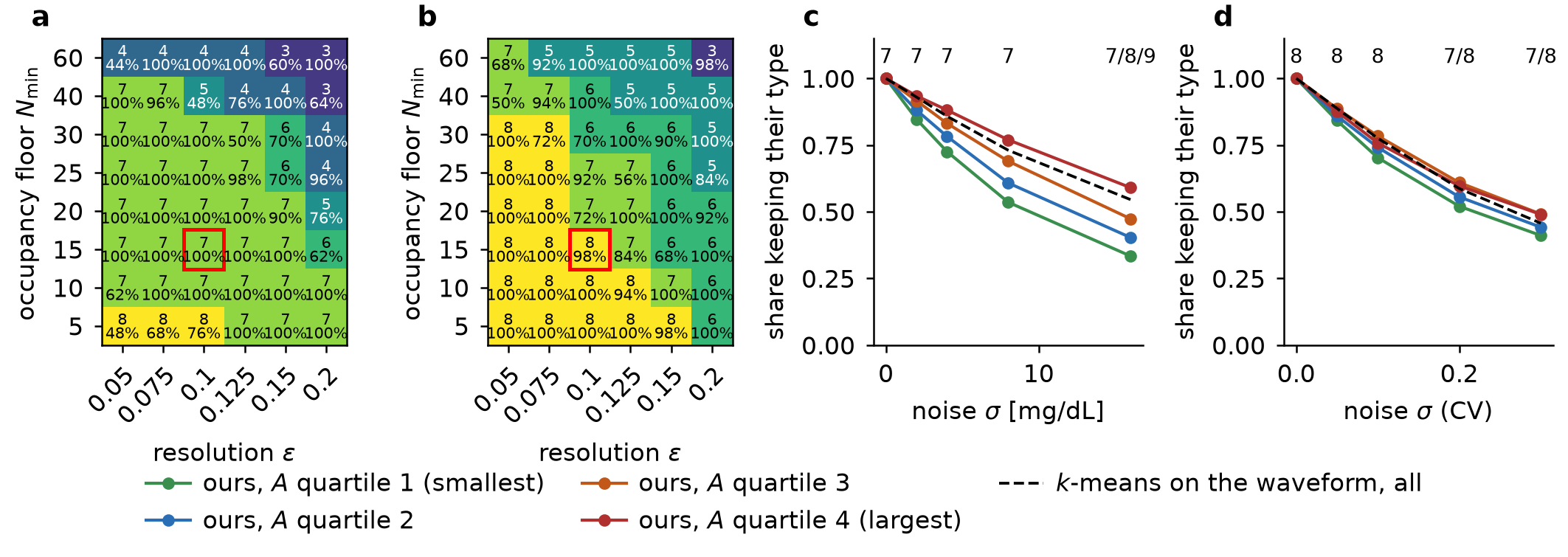}
  \caption{\textbf{Noise reshuffles individual labels before it alters the set of types; the occupancy floor determines which patterns persist once the resolution is set.} \textbf{a--b.} As resolution $\varepsilon$ and occupancy floor $\nmin$ are varied, panels show the number of types (in color and upper number) and the proportion of replicates yielding that count (lower number) for glucose (a) and insulin (b). The red square highlights the reference setting. \textbf{c--d.} For increasing Gaussian noise ($\sigma$), panels show the fraction of curves retaining their reference type by quartile of excursion amplitude $A$ (glucose: additive noise in \si{mg/dL}; insulin: proportional noise). The dashed line shows the same for $k$-means at the same type count, and numbers above each panel indicate the type count from a full refit on the noisy data.}\label{fig:sensitivity}
\end{figure}

To test the robustness of glycemia and insulinemia type identification, we performed a sensitivity analysis by varying both the resolution and the occupancy floor. Glycemia types remain robust across a wide range of $\nmin$ and $\varepsilon$ (Fig.~\ref{fig:sensitivity}a). For insulinemia, eight groups are reliably identified for $\nmin \leq 15$ at $\varepsilon=0.1$ and for $\nmin \leq 30$ at $\varepsilon \leq 0.075$. Changing the weight ratio $w_0/w_1$ (from value-only to difference-only and ratios $1/4$ to $4$), the type counts remain at $7$ and $8$. The main effect is on the assignment of curves compatible with more than one type, with ARI values of $0.72$--$0.88$ for glucose and $0.60$--$0.86$ for insulin compared to the reference. Thus, the occupancy floor sets which patterns are kept, and the distance metric then assigns ambiguous curves.

%Topic: Noise of the size of the repeat-test variability leaves the vocabulary and moves the individual labels, for ours as for k-means.

To assess robustness to noise, we add uncorrelated Gaussian noise to each point on the curves (additive for glycemia, proportional for insulinemia), then re-encode and assign them to the reference types. For glucose, the proportion of curves with unchanged type falls from $0.90$ at $\sigma=2$~\si{mg/dL} to $0.65$ at $8$ and $0.45$ at $16$~\si{mg/dL}, matching the repeat-test scale. This effect is strongest in the quartile with the smallest excursions ($0.33$ versus $0.59$ for the largest at $\sigma=16$~\si{mg/dL}), consistent with the resolution $\varepsilon A$. The $k$-means share drops in parallel, about $0.1$ higher at each $\sigma$ ($0.55$ at $16$~\si{mg/dL}), showing that label instability comes from the shape information, not the encoding. The types themselves remain stable: re-fitting on noisy data yields $7$ glucose types up to $\sigma=8$~\si{mg/dL}, and $8$ in most runs at $16$~\si{mg/dL} (7--9). For insulin, $8$ types are found up to coefficient of variation $0.1$, and $7$ or $8$ beyond that. Proportional noise in insulin affects all quartiles similarly because it scales with amplitude. Thus, at the repeat-test variability of the OGTT, population-level structure is reproducible, but individual labels are not, for both methods. The ratio $\varepsilon A/\sigma$ indicates label reliability; $\varepsilon A$ exceeds $8$~\si{mg/dL} for a quarter of glucose curves and $16$~\si{mg/dL} for less than one percent.

\begin{figure}[!ht]
  \centering
  \includegraphics[width=\linewidth]{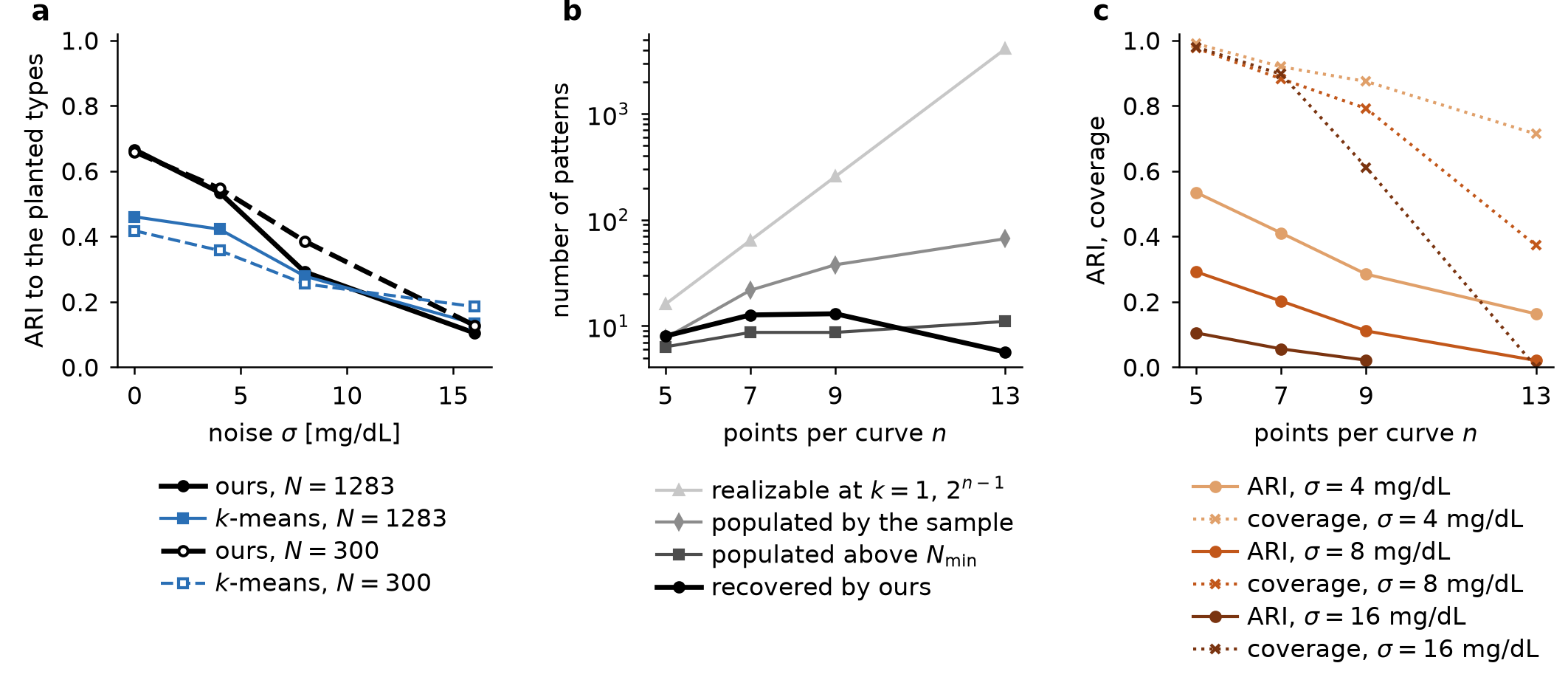}
  \caption{\textbf{The encoding recovers the planted types on synthetic excursions and reveals how the number of time points defines the short-series regime.} \textbf{a.} As noise increases, the panel shows how well the recovered types match the planted types (ARI), for $n=5$ points and two sample sizes; $k$-means at the planted count is shown for comparison. \textbf{b.} As the number of points $n$ increases across a fixed window, the panel traces the total possible patterns ($2^{n-1}$ for $k=1$), those populated in the sample, those above the occupancy floor, and the recovered types ($\sigma=8$~\si{mg/dL}, $N=1283$). \textbf{c.} The final panel shows how agreement and coverage vary with $n$ for three noise levels, using only our encoding. All results are averaged over three seeds.}\label{fig:synthetic}
\end{figure}

%Topic: Synthetic excursions with a planted pattern show what the encoding recovers, and where short stops being short.
Finally, we test whether our method can recover known types from a noisy synthetic dataset (Fig.~\ref{fig:synthetic}). Each curve begins at a basal level after perturbation and follows a gamma-like impulse response; some exhibit a biphasic or late-rebound phase. Most curves return to baseline within the window, while a minority peak beyond it. Levels and amplitudes are sampled from the glucose cohort, so noise in \si{mg/dL} matches the data. The planted type is the sign pattern of the noise-free curve on the grid, matching the method's vocabulary. For $n=5$ and $N=1283$, the recovered types match the planted types at ARI $0.67$ without noise, compared to $0.46$ for $k$-means; agreement holds up to $\sigma=8$~\si{mg/dL}. Noise creates extra patterns above the floor: recovered counts rise from $5$ without noise to $8$ at $\sigma=8$ and $11$ at $\sigma=16$~\si{mg/dL}. With more points on the same window, the number of possible patterns grows as $2^{n-1}$ ($2\cdot3^{n-2}$ at $k=2$, Proposition~\ref{prop:count}), but the number of populated patterns increases roughly linearly ($8$, $22$, $38$, $66$ at $n=5$, $7$, $9$, $13$ for $\sigma=8$~\si{mg/dL}), with $6$--$11$ above the floor. Resolution is the first limiting factor: as $\Delta t$ decreases, the change per step shrinks while noise stays fixed. At $\sigma=8$~\si{mg/dL}, coverage drops from $0.98$ at $n=5$ to $0.79$ at $n=9$ and $0.37$ at $n=13$, and encoding outperforms $k$-means only at $n=5$. At $\sigma=4$~\si{mg/dL}, it leads up to $n=7$; at $n=13$ and $\sigma=16$~\si{mg/dL}, no pattern survives. For this encoding, "short" means the curve's change per grid step still exceeds the noise; in the OGTT window, this holds for $n \leq 7$.


\clearpage
%TC:endignore

\clearpage
\setcounter{figure}{0}
\renewcommand{\thefigure}{S\arabic{figure}}
\section*{Supplemental Material}
\begin{figure}[b]
    \centering
    \includegraphics[width=160mm]{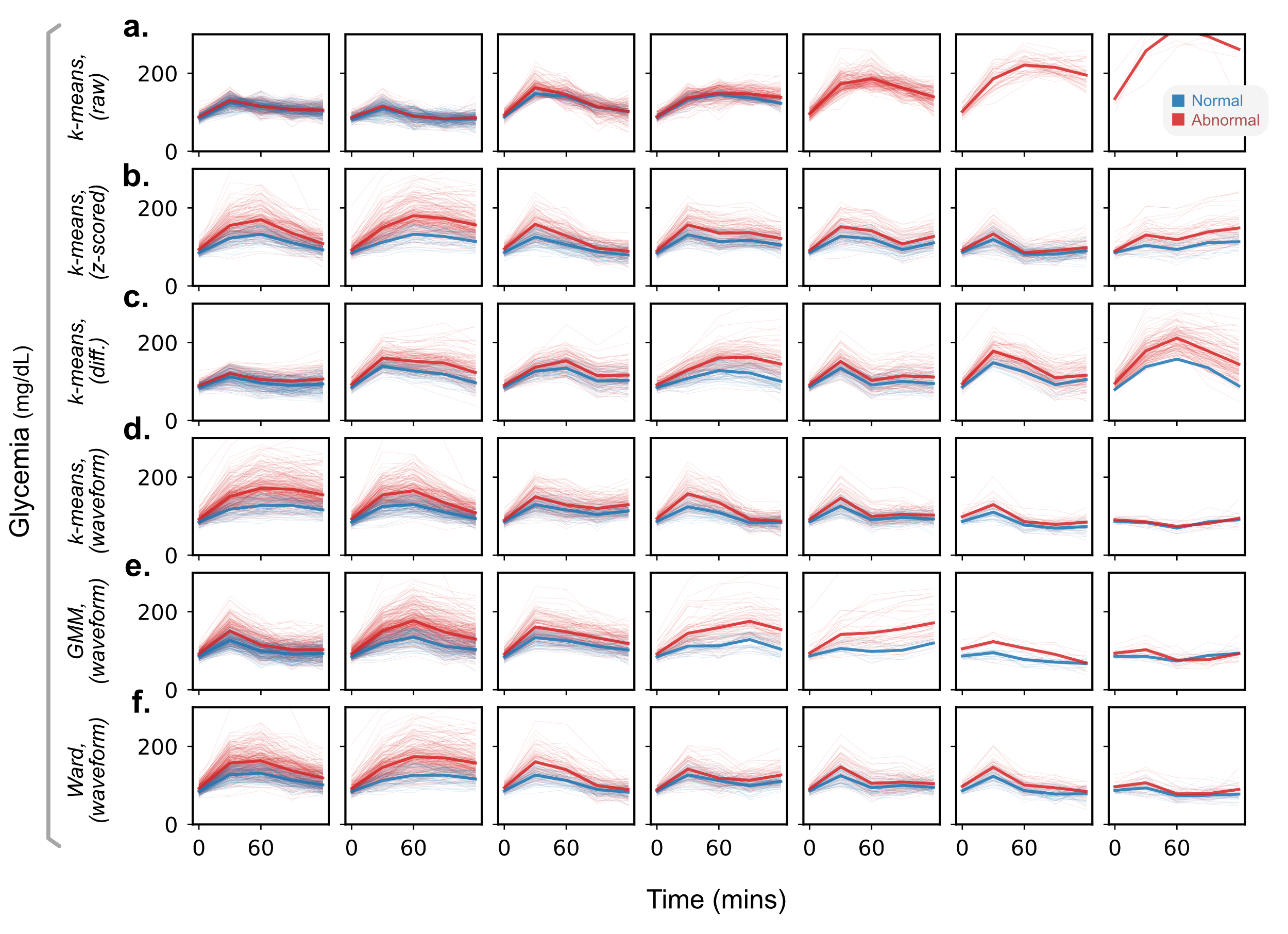}
    \caption{\textbf{At matched count $g=7$, the six classical benchmarked methods cut the cohort mainly along level and excursion, leaving the response shape mixed inside each group.} Columns are each method's seven groups ordered by size, and members are colored by clinical criteria (blue: normal, red: abnormal), with the (clinical) class means in bold.}
    \label{fig:benchG}
\end{figure}
\begin{figure}[tb]
    \centering
    \includegraphics[width=160mm]{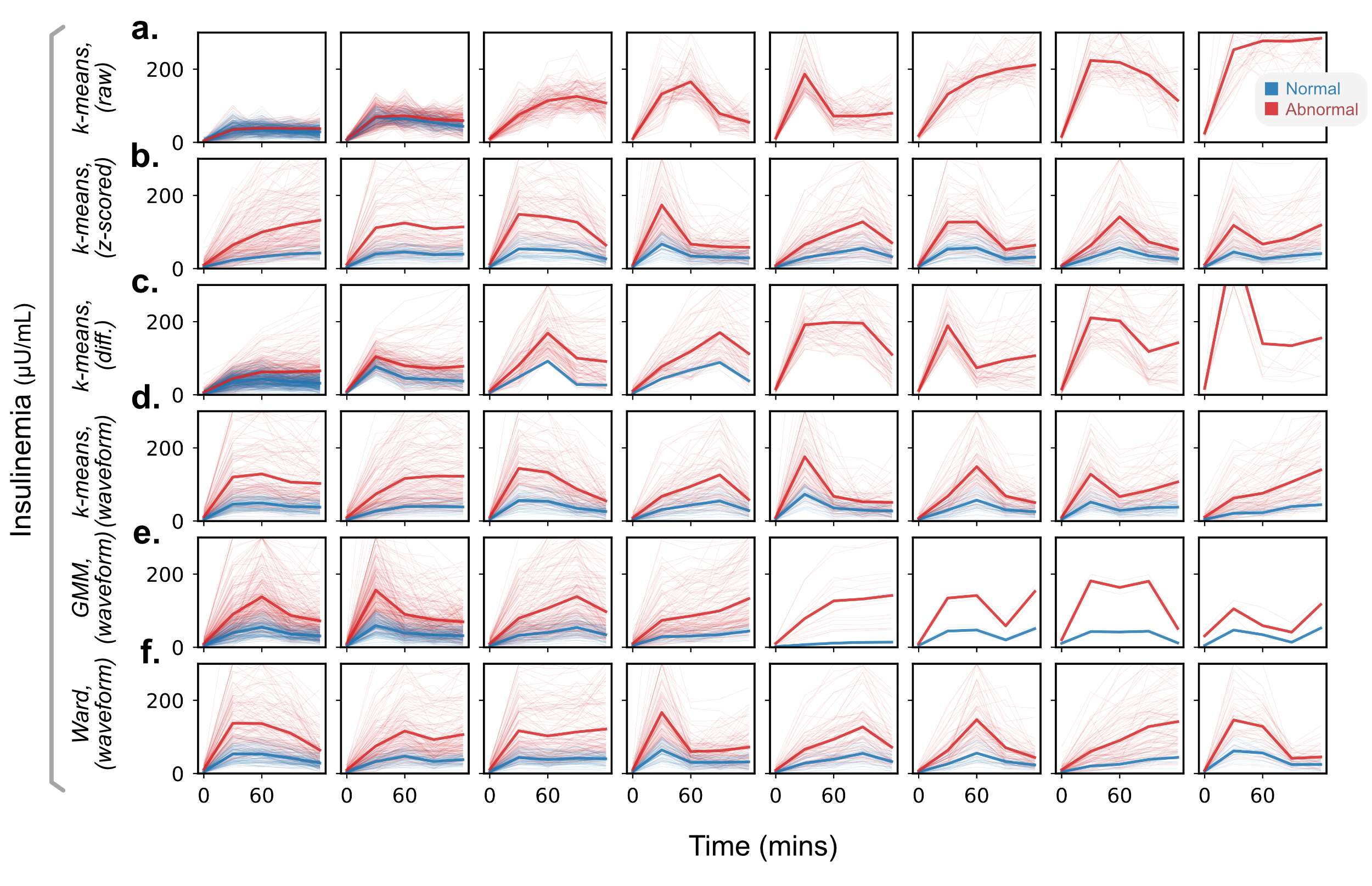}
    \caption{\textbf{The insulin benchmarked methods behave as their glucose counterparts: the groups (now matched at $g=8$) separate by magnitude while each keeps a mixture of shapes.} Columns are each method's eight groups ordered by size, and members are colored by clinical criteria (blue: normal, red: abnormal), with the (clinical) class means in bold.}
    \label{fig:benchI}
\end{figure}
\begin{figure}[bt]
  \centering
  \includegraphics[width=\linewidth]{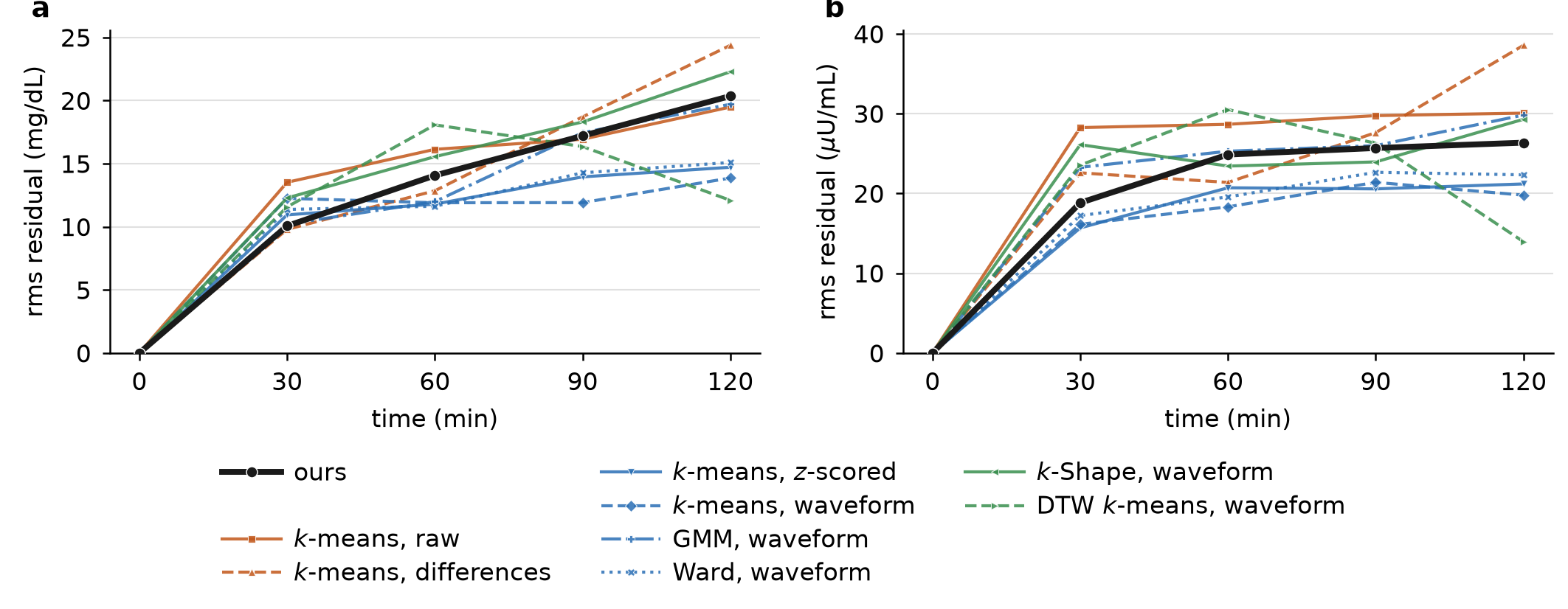}
  \caption{\textbf{The reconstruction error is not spread evenly along the curve.} Root-mean-square residual of the reconstruction of Table~\ref{tab:benchmark} at each point of the protocol grid, in \si{mg/dL} (glucose, \textbf{a}) and \si{\micro U/mL} (insulin, \textbf{b}), at the matched counts $g=7$ and $g=8$. The fasting point is exact for every method. Color separates the partitions that cut the cohort by level (orange) from those that cut it by shape (blue) and the two time-series methods (green); ours is in black.}\label{fig:residual_profile}
\end{figure}

\begin{figure}[tb]
    \centering
    \includegraphics[width=130mm]{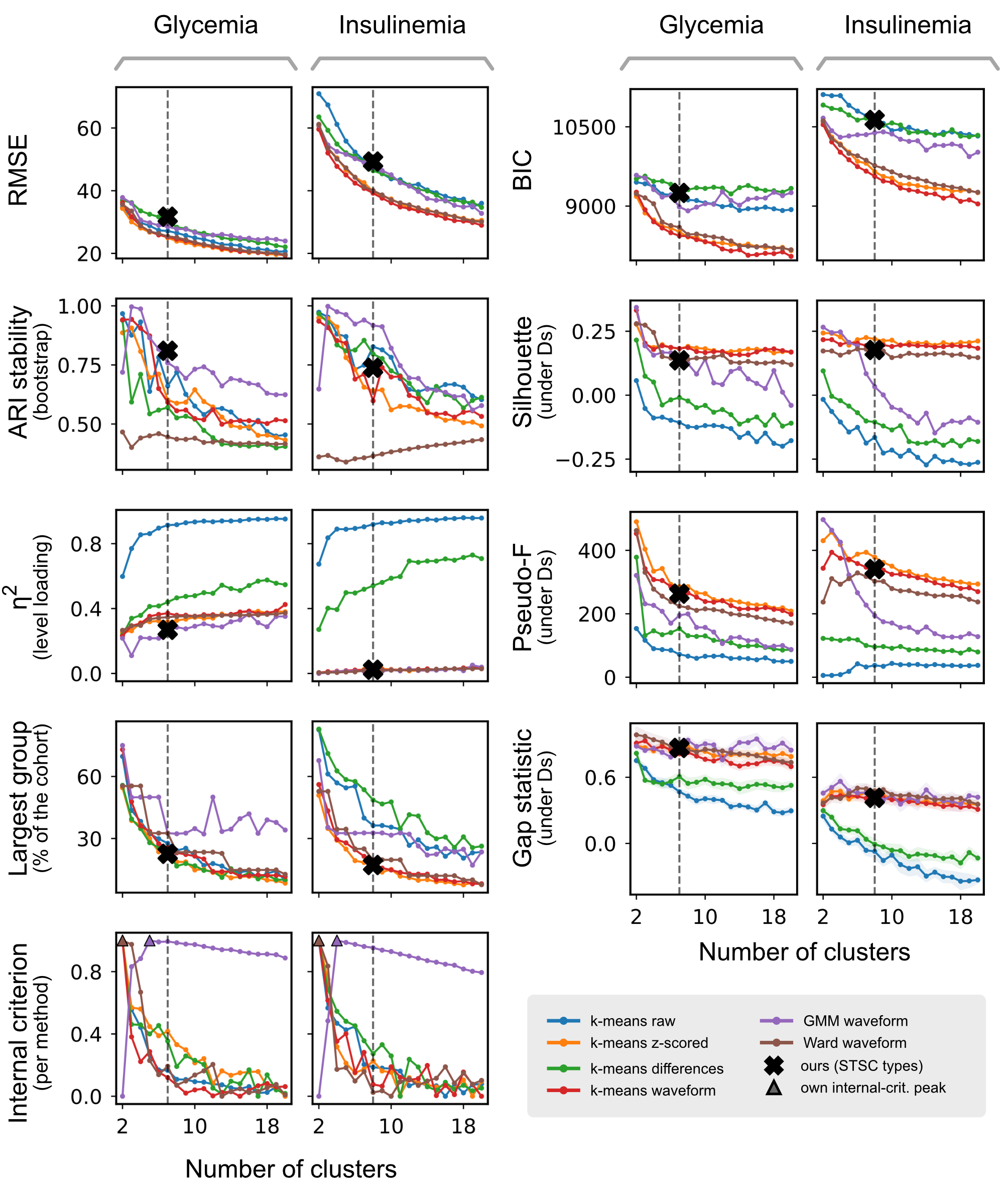}
    \caption{\textbf{No selection criterion available to a benchmarked method points at $g=7$ or $g=8$.} Both blocks sweep $g$ over $2$--$20$ for the six classical methods of Table~\ref{tab:benchmark}. \textbf{a.} Diagnostics independent of $D_s^k(w)$: out-of-sample RMSE, bootstrap ARI stability, $\eta^2$ of the mean level, largest-group share, and each method's native selection criterion (silhouette or BIC for the GMM; min--max normalized per method, gray triangle at its peak); RMSE improves with every added group while stability and the native criteria favor few. \textbf{b.} Diagnostics evaluated under $D_s^k(w)$: BIC-penalized reconstruction, silhouette, pseudo-$F$, and the gap statistic (shaded band, $\pm1$ bootstrap SE); the three validity diagnostics turn over early while the penalized reconstruction keeps improving. The dashed line marks the matched count, and the black $\times$ marks our method, where comparable.}
    \label{fig:groupsweep}
\end{figure}

\end{document}